\documentclass[12pt]{article}
\usepackage{amsmath}
\usepackage{graphicx}

\newsavebox{\boxa}
\sbox{\boxa}{\includegraphics[width=1.6cm]{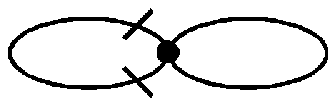}} 
\newlength{\boxal}
\newsavebox{\boxb}
\sbox{\boxb}{\includegraphics[width=1.6cm]{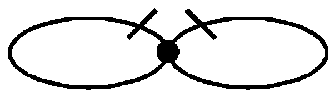}} 
\newlength{\boxbl}
\newsavebox{\boxc}
\sbox{\boxc}{\includegraphics[width=0.6cm]{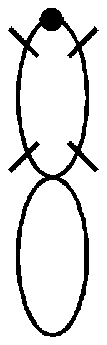}} 
\newlength{\boxcl}
\newsavebox{\boxd}
\sbox{\boxd}{\includegraphics[width=0.6cm]{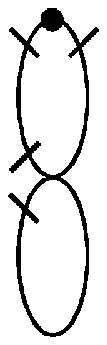}} 
\newlength{\boxdl}
\newsavebox{\boxe}
\sbox{\boxe}{\includegraphics[width=0.6cm]{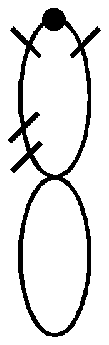}} 
\newlength{\boxel}
\newsavebox{\boxxa}
\sbox{\boxxa}{\includegraphics[width=1.4cm]{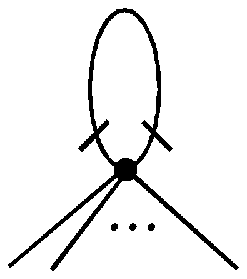}} 
\newlength{\boxxal}
\newsavebox{\boxxb}
\sbox{\boxxb}{\includegraphics[width=1.4cm]{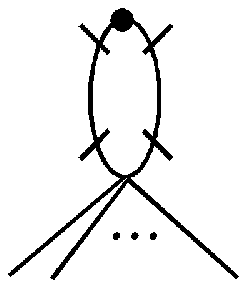}} 
\newlength{\boxxbl}
\newsavebox{\boxxc}
\sbox{\boxxc}{\includegraphics[width=2.5cm]{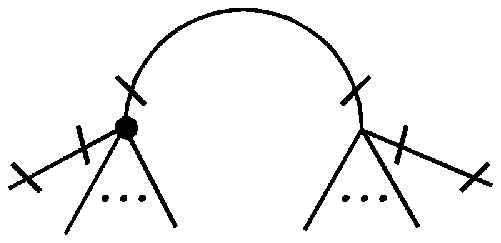}} 
\newlength{\boxxcl}
\newsavebox{\boxxd}
\sbox{\boxxd}{\includegraphics[width=2.5cm]{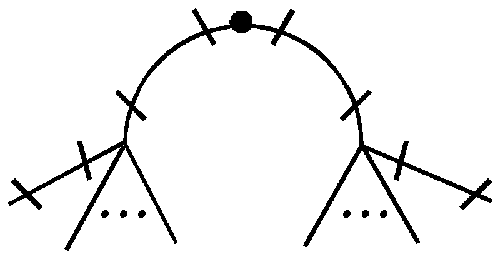}} 
\newlength{\boxxdl}
\newsavebox{\chaina}
\sbox{\chaina}{\includegraphics[width=0.6cm]{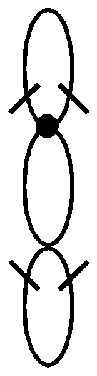}} 
\newlength{\chainal}
\newsavebox{\chainb}
\sbox{\chainb}{\includegraphics[width=0.6cm]{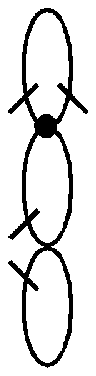}} 
\newlength{\chainbl}
\newsavebox{\chainc}
\sbox{\chainc}{\includegraphics[width=0.6cm]{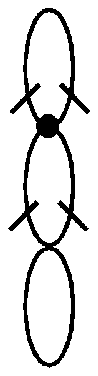}} 
\newlength{\chaincl}
\newsavebox{\chaind}
\sbox{\chaind}{\includegraphics[width=0.6cm]{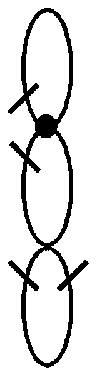}} 
\newlength{\chaindl}
\newsavebox{\chaine}
\sbox{\chaine}{\includegraphics[width=0.6cm]{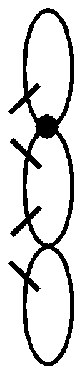}} 
\newlength{\chainel}
\newsavebox{\chainf}
\sbox{\chainf}{\includegraphics[width=0.6cm]{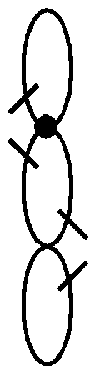}} 
\newlength{\chainfl}
\newsavebox{\chaing}
\sbox{\chaing}{\includegraphics[width=0.6cm]{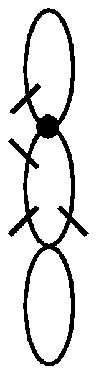}} 
\newlength{\chaingl}
\newsavebox{\chainh}
\sbox{\chainh}{\includegraphics[width=0.6cm]{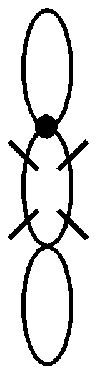}} 
\newlength{\chainhl}
\newsavebox{\chaini}
\sbox{\chaini}{\includegraphics[width=0.6cm]{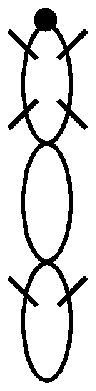}} 
\newlength{\chainil}
\newsavebox{\chainj}
\sbox{\chainj}{\includegraphics[width=0.6cm]{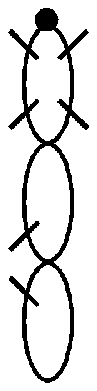}} 
\newlength{\chainjl}
\newsavebox{\chaink}
\sbox{\chaink}{\includegraphics[width=0.6cm]{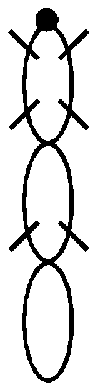}} 
\newlength{\chainkl}
\newsavebox{\chainl}
\sbox{\chainl}{\includegraphics[width=0.6cm]{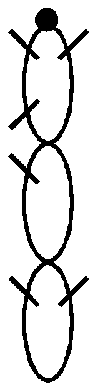}} 
\newlength{\chainll}
\newsavebox{\chainm}
\sbox{\chainm}{\includegraphics[width=0.6cm]{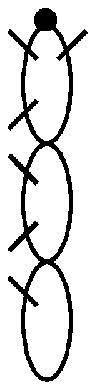}} 
\newlength{\chainml}
\newsavebox{\chainn}
\sbox{\chainn}{\includegraphics[width=0.6cm]{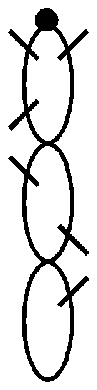}} 
\newlength{\chainnl}
\newsavebox{\chaino}
\sbox{\chaino}{\includegraphics[width=0.6cm]{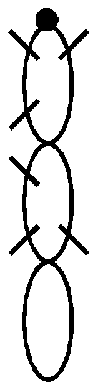}} 
\newlength{\chainol}
\newsavebox{\chainp}
\sbox{\chainp}{\includegraphics[width=0.6cm]{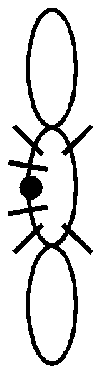}} 
\newlength{\chainpl}
\newsavebox{\chainq}
\sbox{\chainq}{\includegraphics[width=0.6cm]{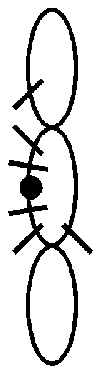}} 
\newlength{\chainql}
\newsavebox{\circlea}
\sbox{\circlea}{\includegraphics[width=1.2cm]{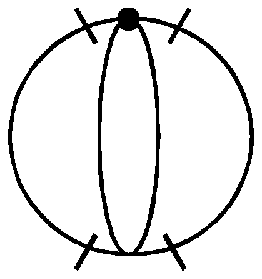}} 
\newlength{\circleal}
\newsavebox{\circleb}
\sbox{\circleb}{\includegraphics[width=1.2cm]{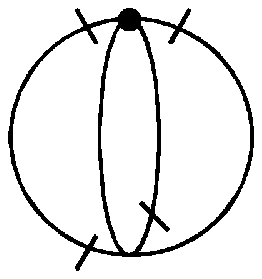}} 
\newlength{\circlebl}
\newsavebox{\circlec}
\sbox{\circlec}{\includegraphics[width=1.2cm]{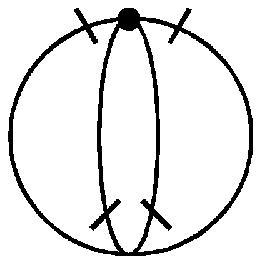}} 
\newlength{\circlecl}
\newsavebox{\circled}
\sbox{\circled}{\includegraphics[width=1.2cm]{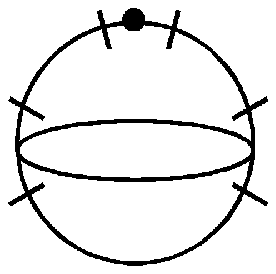}} 
\newlength{\circledl}
\newsavebox{\circlee}
\sbox{\circlee}{\includegraphics[width=1.2cm]{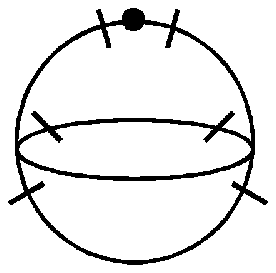}} 
\newlength{\circleel}
\newsavebox{\circlef}
\sbox{\circlef}{\includegraphics[width=1.2cm]{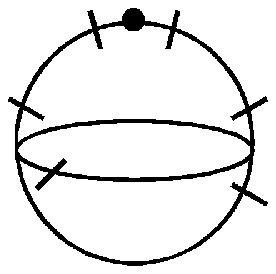}} 
\newlength{\circlefl}
\newsavebox{\circleg}
\sbox{\circleg}{\includegraphics[width=1.2cm]{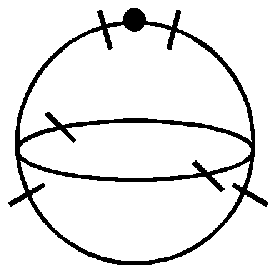}} 
\newlength{\circlegl}
\newsavebox{\circleh}
\sbox{\circleh}{\includegraphics[width=1.2cm]{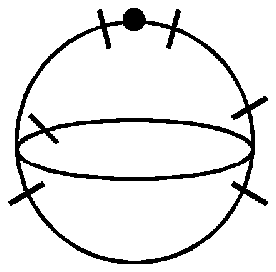}} 
\newlength{\circlehl}
\newsavebox{\circlei}
\sbox{\circlei}{\includegraphics[width=1.2cm]{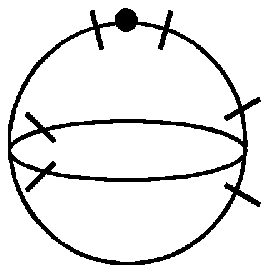}} 
\newlength{\circleil}
\newsavebox{\clovera}
\sbox{\clovera}{\includegraphics[width=1.2cm]{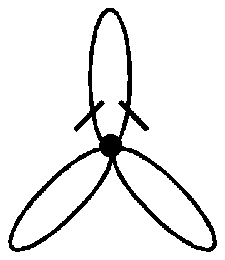}} 
\newlength{\cloveral}
\newsavebox{\cloverb}
\sbox{\cloverb}{\includegraphics[width=1.2cm]{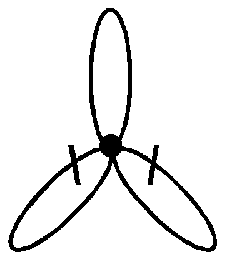}} 
\newlength{\cloverbl}
\newsavebox{\cloverc}
\sbox{\cloverc}{\includegraphics[width=1.2cm]{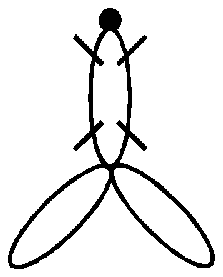}} 
\newlength{\clovercl}
\newsavebox{\cloverd}
\sbox{\cloverd}{\includegraphics[width=1.2cm]{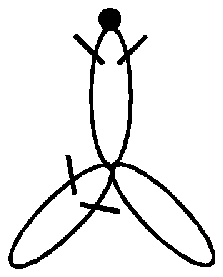}} 
\newlength{\cloverdl}
\newsavebox{\clovere}
\sbox{\clovere}{\includegraphics[width=1.2cm]{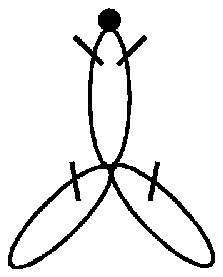}} 
\newlength{\cloverel}
\newsavebox{\cloverf}
\sbox{\cloverf}{\includegraphics[width=1.2cm]{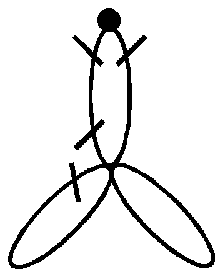}} 
\newlength{\cloverfl}
\newsavebox{\countera}
\sbox{\countera}{\includegraphics[width=0.6cm]{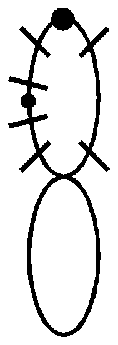}} 
\newlength{\counteral}
\newsavebox{\counterb}
\sbox{\counterb}{\includegraphics[width=0.6cm]{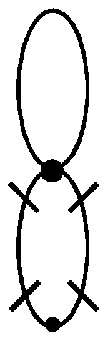}} 
\newlength{\counterbl}
\newsavebox{\counterc}
\sbox{\counterc}{\includegraphics[width=0.6cm]{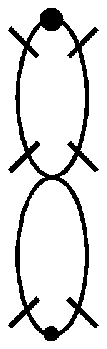}} 
\newlength{\countercl}
\newsavebox{\counterd}
\sbox{\counterd}{\includegraphics[width=1.6cm]{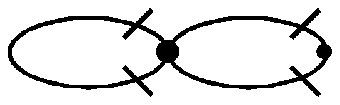}} 
\newlength{\counterdl}
\def\papertitlepage{\baselineskip 3.5ex\thispagestyle{empty}}
\def\preprinumber#1#2{\hfill\begin{minipage}{4.2cm} #1
        \par\noindent #2 \end{minipage}}
\allowdisplaybreaks[3]

\begin{document}

\papertitlepage
\setcounter{page}{0}
\preprinumber{KEK-TH-1492}{}
\baselineskip 0.8cm
\vspace*{2.0cm}

\begin{center}
{\Large\bf Infra-red effects of Non-linear sigma model\\ in de Sitter space}
\end{center}

\begin{center}

Hiroyuki K{\sc itamoto}$^{2)}$
\footnote{E-mail address: kitamoto@post.kek.jp}
and
Yoshihisa K{\sc itazawa}$^{1),2)}$
\footnote{E-mail address: kitazawa@post.kek.jp}\\
\vspace{5mm}
$^{1)}$
{\it KEK Theory Center}\\
{\it Tsukuba, Ibaraki 305-0801, Japan}\\
$^{2)}$
{\it The Graduate University for Advanced Studies (Sokendai)}\\
{\it Department of Particle and Nuclear Physics}\\
{\it Tsukuba, Ibaraki 305-0801, Japan}\\

\end{center}

\vskip 5ex
\baselineskip = 2.5 ex

\begin{center}{\bf Abstract}\end{center}
We extend our investigation on a possible de Sitter symmetry breaking mechanism in non-linear sigma models.
The scale invariance of the quantum fluctuations could make the cosmological constant time dependent
signaling the de Sitter symmetry breaking. 
To understand such a symmetry breaking mechanism, we investigate the energy-momentum tensor.
We show that the leading infra-red logarithms cancel to all orders in perturbation theory in a generic
non-linear sigma model.
When the target space is an $N$ sphere, the de Sitter symmetry is preserved in the large $N$ limit.
For a less symmetric target space, the infra-red logarithms appear at the three loop level.
However there is a counter term to precisely cancel it. 
The leading infra-red logarithms do not cancel for higher derivative interactions.
We investigate such a model in which the infra-red logarithms first appear at the three loop level.
A nonperturbative investigation in the large $N$ limit shows that they eventually grow as large as
the one loop effect. 

\vspace*{\fill}
\noindent
March 2012

\newpage
\section{Introduction}
\setcounter{equation}{0}

It is well known that there is no de Sitter (dS) invariant vacuum 
in a free massless minimally coupled scalar field theory. 
In the exponentially expanding universe, 
the degrees of freedom outside the cosmological horizon increase with cosmic evolution. 
This increase gives rise to a growing time dependence of the propagator \cite{Vilenkin1982,Linde1982,Starobinsky1982}. 
So in some field theoretic models on dS space, 
a certain physical constant may become time dependent through the propagator. 
In particularly, this infra-red (IR) effect may be relevant to resolve the cosmological constant problem
\cite{Woodard1996}.

In order to investigate the IR effects in field theoretic models on dS space, 
we need to employ 
the Schwinger-Keldysh perturbation theory \cite{Schwinger1961,Keldysh1964}. 
The IR effects manifest as the polynomials in the logarithm of the scale factor 
of the universe $\log a(\tau)$ at each order \cite{Weinberg2005}. 
For example in $\lambda\varphi^4$ theory, the leading IR effect
to the potential is the $2n$-th power of the logarithm at the $n$-th order of the coupling $\lambda$
\cite{Woodard2002}. 
These results indicate that the perturbation theory eventually breaks down after a large enough cosmic expansion.
In oder to understand such a situation, we have to investigate the IR effect nonperturbatively. 
Remarkably in the models with interaction potentials, 
the leading IR effects can be evaluated nonperturbatively by the stochastic approach \cite{Starobinsky1994,Woodard2005}. 
However in a general model with derivative interactions, 
we still don't know how to evaluate the nonperturbative infra-red effects . 

As a model with derivative interactions, we have investigated the non-linear sigma model in \cite{Kitamoto2010}. 
It is because the non-linear sigma model contains massless minimally coupled scalar fields.
Furthermore it is exactly solvable in the large $N$ limit on an $S_N$. 
We have evaluated the expectation value of the energy-momentum tensor.
The coefficient in front of the metric tensor $g_{\mu\nu}$ gives the matter contribution to the
cosmological constant.
The IR power counting argument predicts that there are ($n-1$)-th power of $\log a(\tau)$ at the $n$ loop level.
If so, the cosmological constant becomes time dependent at the two loop level.
However we have found that there is a nontrivial cancellation mechanism beyond the power counting arguments.
We have shown that the leading IR effects to the cosmological constant cancel out each other
at the two loop level on an arbitrary target space. For a non-linear sigma model on an $N$ dimensional sphere $S_N$,
we have further shown that the cosmological constant stays the free field value in the large $N$ and
weak coupling limit. It implies that the leading IR logarithms cancel to all orders.

After obtaining these results, there arise two natural questions. 
Firstly we wonder whether the cancellation of the leading IR effects holds beyond the two loop level 
on an arbitrary target space. 
Secondly we would like to know whether the cancellation holds up to the sub-leading IR effect. 
It is our main purpose of this paper to answer these questions. 
In fact we show that the leading IR effects to the cosmological constant cancel out to all orders in perturbation theory.
As for the sub-leading IR effect, it could appear at the three loop level. 
However we find that they again cancel out for a symmetric space. Furthermore we find there is a renormalization scheme to
cancel them for a generic target space.
This cancellation mechanism is specific to the action with two derivatives.
We show that the leading IR effects do not cancel with higher derivative interactions.

The organization of this paper is as follows. 
We introduce a scalar field theory in dS space, in particular, its IR behavior in Section $2$. 
We recall the background field method for the non-linear sigma model in Section $3$.  
In Section $4$, we prove that the leading IR effects to the cosmological constant cancel out each other 
on an arbitrary target space. 
In Section $5$, we evaluate the the sub-leading IR effect to the cosmological constant at the two loop level. 
In Section $6$, we restrict the target space to an $S_N$ and consider the large $N$ limit. 
We find that the effective cosmological constant in this case is time independent to all orders. 
In Section $7$, we investigate whether the effective cosmological constant on an arbitrary target space is time
dependent or not at the three loop level. 
In Section $8$, we evaluate the IR effects in a field theory with a higher derivative interaction term. 
In this model, the leading IR effects to the cosmological constant don't cancel out each other. 
We conclude with discussions in Section $9$. 

\section{Scalar field in the de Sitter space}
\setcounter{equation}{0}

In the Poincar\'{e} coordinate, the metric in de Sitter (dS) space is
\begin{equation}
ds^2=-dt^2+a^2(t)d{\bf x}^2,\hspace{1em}a(t)=e^{Ht}, 
\end{equation}
where the dimension of dS space is taken as $D=4$ and $H$ is the Hubble constant. 
In the conformally flat coordinate,
\begin{equation}
g_{\mu\nu}=a^2(\tau)\eta_{\mu\nu},\hspace{1em}a(\tau)=-\frac{1}{H\tau}. 
\end{equation}
Here the conformal time $\tau (-\infty <\tau < 0)$ is related to the cosmic time $t$ as $\tau\equiv-\frac{1}{H}e^{-Ht}$. 

The quadratic action for a massless scalar field which is minimally coupled to the dS background is
\begin{align}
S_{matter}=\frac{1}{2}\int\sqrt{-g}d^4x\ [-g^{\mu\nu}\partial_\mu\varphi\partial_\nu\varphi].
\end{align}
The positive frequency solution of the equation of motion with respect to this action is
\begin{equation}
\phi_{{\bf p}}(x)=\frac{H\tau}{\sqrt{2p}}(1-i\frac{1}{p\tau})\ e^{-ip\tau+i{\bf p}\cdot{\bf x}},
\end{equation}
where $p=|\bf{p}|$. 
We expand the scalar field as
\begin{equation}
\varphi (x) = \int \frac{d^3p}{(2\pi)^3}\left( a_{{\bf p}}\phi_{{\bf p}}(x)
+ a_{{\bf p}}^{\dagger}\phi_{{\bf p}}^*(x)\right). 
\end{equation}
We consider the Bunch-Davies vacuum $|0\rangle$ which is annihilated by all the annihilation operators 
$\forall a_{{\bf p}}|0\rangle=0$.
The propagator in such a vacuum is 
\begin{align}
\langle\varphi(x)\varphi(x')\rangle
=&\int \frac{d^3p}{(2\pi)^3}\ \phi_{{\bf p}}(x)\phi_{{\bf p}}^*(x')\\
=&\int \frac{d^3p}{(2\pi)^3}\ \frac{H^2\tau\tau'}{2p}(1-i\frac{1}{p\tau})(1+i\frac{1}{p\tau'})
\ e^{-ip(\tau-\tau')+i{\bf p}\cdot({\bf x}-{\bf x}')}.\notag
\end{align}

Let us estimate the magnitude of the quantum fluctuation by taking the coincident limit of the propagator.
It consists of the contributions from inside and outside the cosmological horizon as follows 
\begin{equation}
\langle\varphi(x)\varphi(x)\rangle\sim \int_{P>H}\frac{d^3P}{(2\pi)^3}\frac{1}{2P}+H^2\int_{P<H}\frac{d^3P}{(2\pi)^3}\frac{1}{2P^3}, 
\end{equation}
where $P$ denotes the physical momentum $P\equiv p/a(\tau)=H|\tau|p$.
The ultra-violet (UV) contribution $(P>H)$ is quadratically divergent just like in Minkowski space. 
It can be regularized and renormalized in an identical way.
The logarithmic IR divergence due to the contributions from outside the cosmological horizon $(P<H)$ is specific to dS space.
To regularize this IR divergence, we introduce an IR cut-off $\varepsilon_0$  
which fixes the minimum value of the comoving momentum as in \cite{Lyth2007}. 

With this prescription, more degrees of freedom go out of the cosmological horizon at $P=H$ with cosmic evolution. 
In contrast, the UV cut-off $\Lambda_{UV}$ fixes the maximum value of the physical momentum.
While the degrees of freedom inside the cosmological horizon remains constant, 
the degrees of freedom outside the cosmological horizon increases as time goes on. 
The contribution from outside the cosmological horizon gives a growing time dependence to the propagator 
\begin{align}
\langle\varphi(x)\varphi(x)\rangle&=(\text{UV const})+\frac{H^2}{4\pi^2}\int^H_{\varepsilon_0a^{-1}(\tau)}\frac{dP}{P}\\
&=(\text{UV const})+\frac{H^2}{4\pi^2}\log\big(\frac{H}{\varepsilon_0}a(\tau)\big).\notag
\end{align}
Physically speaking, we consider a situation that a universe with a finite 
spatial extension or a finite region of space starts dS expansion at an initial time $t_i$.
The IR cut-off $\varepsilon_0$ is identified with the initial time $t_i$ as 
\begin{equation}
\log\big(\frac{H}{\varepsilon_0}a(\tau)\big)=\log e^{H(t-t_i)},
\hspace{1em}t_i\equiv\frac{1}{H}\log\frac{\varepsilon_0}{H}. 
\end{equation}
Henceforth we adopt the following setting for simplicity
\begin{equation}
\varepsilon_0=H\Leftrightarrow t_i=0. 
\end{equation}
In this setting, the propagator is 
\begin{equation}
\langle\varphi(x)\varphi(x)\rangle=(\text{UV const})+\frac{H^2}{4\pi^2}\log a(\tau). 
\end{equation}
This time dependence breaks the dS invariance as has been pointed out in \cite{Vilenkin1982,Linde1982,Starobinsky1982}. 

In the subsequent investigations, we evaluate the quantum loop effects. 
For this purpose, we need to regularize the UV divergences of the loop amplitudes. 
In this paper, we adopt the dimensional regularization. 
In $D=4-\varepsilon$, the propagator for the massless and minimally coupled scalar field is expressed as \cite{Miao2008,Miao2010}
\begin{align}
\langle\varphi(x)\varphi(x')\rangle=A(y)+B\log(a(\tau)a(\tau')), 
\label{G}\end{align}
where $y$ is a dS invariant distance function
\begin{equation}
y\equiv\frac{-(\tau-\tau')^2+({\bf x}-{\bf x}')^2}{\tau\tau'}. 
\label{y}\end{equation}
$A(y), B, \delta$ are defined as follows: 
\begin{align}
&A(y)=\frac{H^{D-2}}{(4\pi)^\frac{D}{2}}\Big\{
\Gamma(\frac{D}{2}-1)\big(\frac{4}{y}\big)^{\frac{D}{2}-1}
+\frac{\Gamma(\frac{D}{2}+1)}{\frac{D}{2}-2}\big(\frac{y}{4}\big)^{2-\frac{D}{2}}
+\frac{\Gamma(D-1)}{\Gamma(\frac{D}{2})}\delta\label{Gdetail}\\
&\hspace{5em}+\sum^{\infty}_{n=1}\big[\frac{\Gamma(D-1+n)}{n\Gamma(\frac{D}{2}+n)}
\big(\frac{y}{4}\big)^n
-\frac{\Gamma(\frac{D}{2}+1+n)}{(2-\frac{D}{2}+n)(n+1)!}\big(\frac{y}{4}\big)^{n+2-\frac{D}{2}}\big]\Big\},\notag\\
&B=\frac{H^{D-2}}{(4\pi)^\frac{D}{2}}
\frac{\Gamma(D-1)}{\Gamma(\frac{D}{2})},\notag\\
&\delta=-\psi(1-\frac{D}{2})+\psi(\frac{D-1}{2})+\psi(D-1)+\psi(1),\hspace{1em}\psi(z)\equiv\Gamma'(z)/\Gamma(z). \notag 
\end{align}
Although the dimensional regularization doesn't break the dS symmetry, 
the IR cut-off breaks it and induces the $\log(a(\tau)a(\tau'))$ term. 

We investigate the expectation value of the energy-momentum tensor
as we are interested in how the IR logarithms contribute to the cosmological constant. 
The energy-momentum tensor appears on the right-hand side of the Einstein equation
\begin{align}
R_{\mu\nu}-\frac{1}{2}g_{\mu\nu}R+\Lambda g_{\mu\nu}&=\kappa T_{\mu\nu},\hspace{1em}\kappa=8\pi G,\label{Eeq}\\
T_{\mu\nu}&\equiv\frac{-2}{\sqrt{-g}}\frac{\delta S_{matter}}{\delta g^{\mu\nu}}, \notag
\end{align}
where $\Lambda$ is the cosmological constant and $G$ is the Newton's constant. 
As far as the dS symmetry is preserved, 
the vacuum expectation value(vev) of the energy-momentum tensor is proportional to $g_{\mu\nu}$
with a constant coefficient
\begin{align}
\langle T_{\mu\nu}\rangle=g_{\mu\nu}T. 
\end{align}
If the coefficient of $g_{\mu\nu}$ becomes time dependent,
the de Sitter symmetry is broken down to the spatial rotation and spatial translation. 
The term which is proportional to $\delta_\mu^{\ 0}\delta_\nu^{\ 0}$ emerges to preserve the covariant conservation 
low of the energy momentum tensor
\begin{align}
\langle T_{\mu\nu}\rangle=&g_{\mu\nu}T(\tau)+a^2(\tau)\delta_\mu^{\ 0}\delta_\nu^{\ 0}U(\tau),\label{conservation}\\ 
&U(\tau)=\tau^3\int d\tau\ \tau^{-3}\frac{d}{d\tau}T(\tau)
\ \Leftarrow\ \nabla_\mu\langle T^\mu_{\ \nu}\rangle=0.\notag
\end{align}
Since the time dependence is caused by the IR logarithms, $T$ is logarithmically larger than $U$.
It is in this sense that the matter quantum IR effect could induce the time dependent effective cosmological constant
\begin{align}
\Lambda_{eff}=\Lambda-\kappa T(\tau). 
\end{align}

In the free field theory, the vev of the energy-momentum tensor is
\begin{align}
\langle T_{\mu\nu}\rangle= (\delta_\mu^{\ \rho}\delta_\nu^{\ \sigma}-\frac{1}{2}g_{\mu\nu}g^{\rho\sigma})
\langle\partial_\rho\varphi\partial_\sigma\varphi\rangle. 
\end{align}
From (\ref{G})-(\ref{Gdetail}), we find that the contribution from the free field is time independent
\begin{align}
\langle T_{\mu\nu}\rangle= \frac{3H^4}{32\pi^2}g_{\mu\nu},\hspace{1em}
\Lambda_{eff}=\Lambda-\kappa \frac{3H^4}{32\pi^2}. 
\label{freeT}\end{align}
In this paper, we work with the Poincar\'{e} coordinate. 
The propagator and the energy-momentum tensor for a free field is investigated 
by using the global coordinate in \cite{Allen1987,Folacci1991}. 
The result is a little different from (\ref{freeT}). 
However the difference rapidly vanishes at late times with the spatial expansion. 
So we believe the Poincar\'{e} coordinate is sufficient to investigate the IR effects which grow with time.
It is necessary that there exist interaction terms which contain undifferentiated scalar fields to identify the IR logarithms. Non-linear sigma models satisfy this necessary condition.
In the subsequent sections, we investigate the IR effects in the non-linear sigma model 
and a model with a higher derivative interaction. 

Before investigating the interaction effects, we refer to the conformal anomaly. 
The conformal anomaly also contributes to the vev of the energy-momentum tensor \cite{BD}.
In the case of the minimally coupled scalar field in dS space, 
it leads to the following energy momentum tensor in addition
\begin{equation}
\langle T_{\mu\nu}\rangle=\frac{29H^4}{15\cdot 64\pi^2}g_{\mu\nu}. 
\end{equation}
This contribution has no time dependence because the conformal anomaly is the UV effect. 
In this paper, we focus on the time dependence of the effective cosmological constant induced by IR quantum effects.

\section{Non-linear sigma model}
\setcounter{equation}{0}

In this paper, we investigate the IR effects of the non-linear sigma model in dS space. 
There are two reasons why we are interested in the non-linear sigma model. 
Firstly the non-linear sigma model contains 
massless and minimally coupled scalar fields due to the reparametrization invariance of the target space. 
Secondly we can investigate nonperturbative effects as it becomes exactly solvable in the large $N$ limit. 

The action of the non-linear sigma model is
\begin{equation}
S_{matter}=\frac{1}{2g^2}\int\sqrt{-g}d^4x\ G_{ij}(\varphi)(-g^{\mu\nu}\partial_\mu\varphi^i\partial_\nu\varphi^j), 
\end{equation}
where $g_{\mu\nu}$ is the metric of the dS space, $g^2$ is the coupling constant 
and $G_{ij}(i=1\cdots N)$ is the metric of the target space. 
The reparaterization invariance of the target space is the important symmetry of the non-linear sigma model
as it follows from the consistency as a quantum theory. The dimensional regularization respects this important
symmetry.
We adopt the background field method which is manifestly covariant.
The action is expanded as follows \cite{AlvarezGaume1981}
\begin{align}
S_{matter}=&\frac{-1}{2g^2}\int\sqrt{-g}d^4x\ 
\big[G_{ij}(\bar{\varphi})g^{\mu\nu}\partial_\mu\bar{\varphi}^i\partial_\nu\bar{\varphi}^j
-R_{cidj}(\bar{\varphi})\xi^c\xi^d g^{\mu\nu}\partial_\mu\bar{\varphi}^i\partial_\nu\bar{\varphi}^j\\
&+(-\frac{1}{12}D_eD_fR_{cidj}(\bar{\varphi})+\frac{1}{3}R^g_{\ cad}R_{gebf}(\bar{\varphi}))\xi^c\xi^d\xi^e\xi^fg^{\mu\nu}\partial_\mu\bar{\varphi}^i\partial_\nu\bar{\varphi}^j\notag\\
&-\frac{4}{3}R_{cidb}(\bar{\varphi})\xi^c\xi^d g^{\mu\nu}(D_\mu\xi)^b\partial_\nu\bar{\varphi}^i\notag\\
&-\frac{1}{2}D_eR_{cidb}(\bar{\varphi})\xi^c\xi^d\xi^e g^{\mu\nu}(D_\mu\xi)^b\partial_\nu\bar{\varphi}^i\notag\\
&+g^{\mu\nu}(D_\mu\xi)^a(D_\nu\xi)^a-\frac{1}{3}R_{cadb}(\bar{\varphi})\xi^c\xi^d g^{\mu\nu}(D_\mu\xi)^a(D_\nu\xi)^b\notag\\
&-\frac{1}{6}D_eR_{cadb}(\bar{\varphi})\xi^c\xi^d\xi^eg^{\mu\nu}(D_\mu\xi)^a(D_\nu\xi)^b\notag\\
&+(-\frac{1}{20}D_eD_fR_{cadb}(\bar{\varphi})+\frac{2}{45}R^g_{\ cad}R_{gebf}(\bar{\varphi}))\xi^c\xi^d\xi^e\xi^fg^{\mu\nu}(D_\mu\xi)^a(D_\nu\xi)^b+\cdots\big], \notag
\end{align}
where $\bar{\varphi}^i$ are the background fields, $\xi^i$ are the quantum fluctuations. 
Here $R_{ikjl}$ is the Riemann tensor
\footnote{Our convention is $R^i_{\ jkl}=\partial_k\Gamma^i_{\ jl} - \partial_l\Gamma^i_{\ jk} +\cdots$ and
$R_{ij}=R^k_{\ ikj}$.}
 and the covariant derivative are
\begin{equation}
D_\mu\xi^i=\partial_\mu\xi^i+\Gamma^i_{\ jk}\partial_\mu\bar{\varphi}^j\xi^k. 
\end{equation}
By using the vielbein $e_i^{\ a}$, we can work in the flat tangential space $E_{N}$ instead of the target space 
\begin{equation}
\xi^a=e_i^{\ a}\xi^i,\hspace{1em}(D_\mu\xi)^a=\partial_\mu\xi^a+\omega_i^{\ ab}\partial_\mu\bar{\varphi}^i\xi^b, 
\end{equation}
where $\omega_i^{\ ab}$ is the spin connection. 
Henceforth we rescale the quantum fluctuations $\xi^a/g\to\xi^a$ for convenience. 

Since we are interested in the contribution to the cosmological constant, we can set the background fields $\bar{\varphi}^i$ zero.  
The vev of the energy-momentum tensor is
\begin{align}
\langle T_{\mu\nu}\rangle=&\ (\delta_\mu^{\ \rho}\delta_\nu^{\ \sigma}-\frac{1}{2}g_{\mu\nu}g^{\rho\sigma})\times\label{NLEMT}\\
&\langle \partial_\rho\xi^a\partial_\sigma\xi^a-\frac{g^2}{3}R_{cadb}\xi^c\xi^d\partial_\rho\xi^a\partial_\sigma\xi^b
-\frac{g^3}{6}D_eR_{cadb}\xi^c\xi^d\xi^e\partial_\rho\xi^a\partial_\sigma\xi^b\notag\\
&+(-\frac{g^4}{20}D_eD_fR_{cadb}+\frac{2g^4}{45}R^g_{\ cad}R_{gebf})\xi^c\xi^d\xi^e\xi^f\partial_\rho\xi^a\partial_\sigma\xi^b+\cdots\rangle. \notag
\end{align}
A propagator left intact by differential operators $\langle\xi(x)\xi(x')\rangle$ can induce a single IR logarithm. 
The power counting procedure for the leading IR logarithms in the expectation value
of the energy-momentum tensor is explained in Appendix A.
The conclusion is that the maximum time dependence of the energy-momentum tensor 
at the $n$-th loop level is $\log^{n-1} a(\tau)$. 
We call it the leading IR effect or the leading IR logarithms. 
It also predicts the $\log^{n-2} a(\tau)$ factor as the sub-leading effect. 
We investigate the leading IR effect in Section 4 and the sub-leading IR effect in Section 5 and 7. 

\section{Cancellation of the leading IR effects}
\setcounter{equation}{0}

In \cite{Kitamoto2010}, we have shown that the leading IR effects to the cosmological constant cancel out each other 
at the two loop level on an arbitrary target space. 
Furthermore we have shown such a cancellation to all orders in the large $N$ limit on an $S_N$
as explained below. 
The power counting of the IR logarithms indicates that even if $g^2H^2\ll 1$, 
the perturbation theory breaks down after $g^2H^2\log a(\tau)\sim 1$. 
In fact we can show that the growth of the IR logarithms stops after that
as the scalar fields become massive.
In such a situation, we need to resum these large logarithms. 
If the leading IR logarithms of the energy-momentum tensor cancel in the nonlinear sigma model, the
matter contribution to the cosmological constant is estimated of order $g^2H^2$.
In fact such an estimate is consistent with the large $N$ limit as we have found the identical result with
the free field theory in the weak coupling limit.
So we are interested in whether the cancellation of the leading IR effects holds to all orders 
on an arbitrary target space. 
In this section, we show that such a cancellation indeed takes place.

We have shown the cancellation of the IR logarithms of the energy-momentum tensor up to the two loop level
in an explicit calculation \cite{Kitamoto2010}. 
R.P. Woodard pointed out to us that such a cancellation can be confirmed by a partial integration method. 
We demonstrate this procedure in what follows as it can be extended beyond the two loop level. 

The contributions to the energy-momentum tensor at two loop level consist of the following two terms 
\begin{align}
&\ \langle\partial_\rho\xi^a\partial_\sigma\xi^a\rangle|_{g^2}\label{first}\\
=&\ \int \sqrt{-g'}d^Dx'\ 
\big\{i\frac{g^2}{3}RG^{++}(x',x')-i(\delta\beta+2\delta\gamma)R\big\}\notag\\
&\times g^{\alpha\beta}(\tau')
\big[\partial_\rho\partial_\alpha' G^{++}(x,x')\partial_\sigma\partial_\beta' G^{++}(x,x')
-\partial_\rho\partial_\alpha' G^{+-}(x,x')\partial_\sigma\partial_\beta' G^{+-}(x,x')\big]\notag\\
&+\int \sqrt{-g'}d^Dx'\ \ 
i\frac{g^2}{3}R\lim_{x''\to x'}\partial_\alpha'\partial_\beta'' G^{++}(x',x'')\notag\\
&\times g^{\alpha\beta}(\tau')
\big[\partial_\rho G^{++}(x,x')\partial_\sigma G^{++}(x,x')
-\partial_\rho G^{+-}(x,x')\partial_\sigma G^{+-}(x,x')\big]\notag\\
&-\int \sqrt{-g'}d^Dx'\ \ 
i\frac{g^2}{6}R\partial_\alpha' G^{++}(x',x')\notag\\
&\times g^{\alpha\beta}(\tau')\partial_\beta'
\big[\partial_\rho G^{++}(x,x')\partial_\sigma G^{++}(x,x')
-\partial_\rho G^{+-}(x,x')\partial_\sigma G^{+-}(x,x')\big], \notag
\end{align}
\begin{align}
&-\frac{g^2}{3}R_{cadb}\langle\xi^c\xi^d\partial_\rho\xi^a\partial_\sigma\xi^b\rangle|_{g^0}
+(\delta\beta+2\delta\gamma)R_{ab}\langle\partial_\rho\xi^a\partial_\sigma\xi^b\rangle|_{g^0}\label{second}\\
=&-\big\{\frac{g^2}{3}RG^{++}(x,x)-(\delta\beta+2\delta\gamma)R\big\}
\lim_{x'\to x}\partial_\rho\partial_\sigma' G^{++}(x,x')\notag\\
&+\frac{g^2}{12}R\partial_\rho G^{++}(x,x)\partial_\sigma G^{++}(x,x). \notag
\end{align}
We have introduced the following counter term to renormalize the UV divergent coefficient 
of $R_{ij}(\bar{\varphi})g^{\mu\nu}\partial_\mu\bar{\varphi}^i\partial_\nu\bar{\varphi}^j$
\begin{align}
-\frac{\delta\beta}{2g^2}R_{ij}(\varphi)g^{\mu\nu}\partial_\mu\varphi^i\partial_\nu\varphi^j,\hspace{1em} 
\delta\beta=g^2\frac{H^{D-2}}{(4\pi)^{\frac{D}{2}}}\frac{\Gamma(D-1)}{\Gamma(\frac{D}{2})}\delta.
\label{beta}\end{align}
We have also renormalized the quantum fluctuations
\begin{align}
\xi^i\to\xi^i+\delta\gamma R^i_{\ j}(\bar{\varphi})\xi^j. 
\label{gamma}\end{align}
Here we set $\delta\gamma$ as follows to renormalize the UV divergence of the kinetic term for the quantum fields
\begin{align}
\delta\beta+2\delta\gamma
=\frac{g^2}{3}\frac{H^{D-2}}{(4\pi)^{\frac{D}{2}}}\frac{\Gamma(D-1)}{\Gamma(\frac{D}{2})}\delta. 
\label{counter}\end{align}
We use the Schwinger-Keldysh formalism to evaluate these expectation values \cite{Schwinger1961,Keldysh1964}
\begin{equation}
G^{++}(x,x')\equiv\langle T\xi_I(x)\xi_I(x')\rangle,\hspace{1em}
G^{+-}(x,x')\equiv\langle \xi_I(x')\xi_I(x)\rangle, 
\end{equation}
where $T$ denotes the time ordering and $\xi_I$ denotes the fields in the interaction picture. 

Here we focus on the leading IR effects 
and retain the parts which can induce a single logarithm $\log a(\tau)$
\begin{align}
&\ \langle\partial_\rho\xi^a\partial_\sigma\xi^a\rangle|_{g^2}\label{first1}\\
\simeq&\ \int \sqrt{-g'}d^Dx'\ 
\big\{i\frac{g^2}{3}RG^{++}(x',x')-i(\delta\beta+2\delta\gamma)R\big\}\notag\\
&\times g^{\alpha\beta}(\tau')
\big[\partial_\rho\partial_\alpha' G^{++}(x,x')\partial_\sigma\partial_\beta' G^{++}(x,x')
-\partial_\rho\partial_\alpha' G^{+-}(x,x')\partial_\sigma\partial_\beta' G^{+-}(x,x')\big],\notag
\end{align}
\begin{align}
&-\frac{g^2}{3}R_{cadb}\langle\xi^c\xi^d\partial_\rho\xi^a\partial_\sigma\xi^b\rangle|_{g^0}
+(\delta\beta+2\delta\gamma)R\langle\partial_\rho\xi^a\partial_\sigma\xi^a\rangle|_{g^0}\label{second1}\\
\simeq&-\big\{\frac{g^2}{3}RG^{++}(x,x)-(\delta\beta+2\delta\gamma)R\big\}
\lim_{x'\to x}\partial_\rho\partial_\sigma' G^{++}(x,x'). \notag
\end{align}
As pointed out in \cite{Woodard2005,Prokopec2006}, 
the partial integration is very useful to evaluate the time dependent contributions
in the diagrams with derivative interactions.   
By using the partial integration, the contribution from the "propagator" term is written as
\begin{align}
&\ \langle\partial_\rho\xi^a\partial_\sigma\xi^a\rangle|_{g^2}\label{first2}\\
\simeq&\ \int d^Dx'\ \partial_\alpha'
\big\{i\frac{g^2}{3}RG^{++}(x',x')-i(\delta\beta+2\delta\gamma)R\big\}\notag\\
&\times \sqrt{-g'}g^{\alpha\beta}(\tau')
\big[\partial_\rho G^{++}(x,x')\partial_\sigma\partial_\beta' G^{++}(x,x')
-\partial_\rho G^{+-}(x,x')\partial_\sigma\partial_\beta' G^{+-}(x,x')\big]\notag\\
&-\int d^Dx'\ \big\{i\frac{g^2}{3}RG^{++}(x',x')-i(\delta\beta+2\delta\gamma)R\big\}\notag\\
&\times \big[\partial_\rho G^{++}(x,x')\partial_\sigma\sqrt{-g'}\nabla'^2G^{++}(x,x')
-\partial_\rho G^{+-}(x,x')\partial_\sigma\sqrt{-g'}\nabla'^2G^{+-}(x,x')\big], \notag
\end{align}
where $\nabla^2=\frac{1}{\sqrt{-g}}\partial_\mu(\sqrt{-g}g^{\mu\nu}\partial_\nu)$. 
Note that the surface term is zero because $\tau\to 0$ is outside the past light corn 
and $\log a(\tau)=0$ at $\tau=-\frac{1}{H}$. 
The first term doesn't induce a single logarithm and so we neglect it. 
By using the following identities
\begin{align}
\sqrt{-g'}\nabla'^2G^{++}(x,x')=i\delta^{(D)}(x-x'),\hspace{1em}\sqrt{-g'}\nabla'^2G^{+-}(x,x')=0, 
\label{nabla}\end{align}
the "propagator" term is
\begin{align}
&\ \langle\partial_\rho\xi^a\partial_\sigma\xi^a\rangle|_{g^2}\label{first3}\\
\simeq&\int d^Dx'\ \big\{\frac{g^2}{3}RG^{++}(x',x')-(\delta\beta+2\delta\gamma)R\big\}
\times \partial_\rho G^{++}(x,x')\partial_\sigma \delta^{(D)}(x-x')\notag\\
=&\big\{\frac{g^2}{3}RG^{++}(x,x)-(\delta\beta+2\delta\gamma)R\big\}
\lim_{x'\to x}\partial_\rho\partial_\sigma' G^{++}(x,x')\notag\\
&+\int d^Dx'\ \partial_\sigma'\big\{\frac{g^2}{3}RG^{++}(x',x')-(\delta\beta+2\delta\gamma)R\big\}
\partial_\rho G^{++}(x,x')\notag\\
\simeq&\big\{\frac{g^2}{3}RG^{++}(x,x)-(\delta\beta+2\delta\gamma)R\big\}
\lim_{x'\to x}\partial_\rho\partial_\sigma' G^{++}(x,x'). \notag
\end{align}
Here we have used the partial integration and neglected the term which doesn't induce a single logarithm. 
From (\ref{second1}) and (\ref{first3}), 
the contributions from the "vertex" term and the "propagator" term cancel out each other up to the leading IR effect 
\begin{align}
\langle\partial_\rho\xi^a\partial_\sigma\xi^a\rangle|_{g^2}
-\frac{g^2}{3}R_{cadb}\langle\xi^c\xi^d\partial_\rho\xi^a\partial_\sigma\xi^b\rangle|_{g^0}
+(\delta\beta+2\delta\gamma)R_{ab}\langle\partial_\rho\xi^a\partial_\sigma\xi^b\rangle|_{g^0}\simeq 0.
\end{align}

The above prescription is easily understood by using the Feynman diagrams. 
The leading IR contributions from the "propagator" term and the "vertex" term 
are represented by the following diagrams
\begin{align}
\langle\partial_\rho\xi^a\partial_\sigma\xi^a\rangle|_{g^2}
\simeq\parbox{\boxcl}{\usebox{\boxc}},
\end{align}
\begin{align}
-\frac{g^2}{3}R_{cadb}\langle\xi^c\xi^d\partial_\rho\xi^a\partial_\sigma\xi^b\rangle|_{g^0}
+(\delta\beta+2\delta\gamma)R\langle\partial_\rho\xi^a\partial_\sigma\xi^a\rangle|_{g^0}
\simeq\parbox{\boxal}{\usebox{\boxa}},
\end{align}
where the dot denotes the location of the energy-momentum tensor $x$. 
The short line segments on the propagator denote the differential operators. 
By using the partial integration, the "propagator" term is
\begin{align}
\parbox{\boxcl}{\usebox{\boxc}}=2\ \parbox{\boxdl}{\usebox{\boxd}}-\parbox{\boxel}{\usebox{\boxe}}. 
\end{align}
We neglect the first diagram because it doesn't induce a single logarithm
\begin{align}
\parbox{\boxcl}{\usebox{\boxc}}\simeq-\parbox{\boxel}{\usebox{\boxe}}.
\label{neglect}\end{align}
Here the double line segments denote $\sqrt{-g'}\nabla'^2$. 
By using (\ref{nabla}) and the partial integration, 
\begin{align}
\parbox{\boxcl}{\usebox{\boxc}}=-2\parbox{\boxbl}{\usebox{\boxb}}-\parbox{\boxal}{\usebox{\boxa}}\simeq-\parbox{\boxal}{\usebox{\boxa}}.
\end{align}
In the last process, we neglected the first diagram since it doesn't induce a IR logarithm. 
As a result, the "propagator" term cancels out the "vertex" term up to the leading IR effect.  

The diagramatic investigation is useful beyond the two loop level. 
We can indeed confirm that the leading IR effects cancel 
between the "propagator" terms and the "vertex" terms.
Let us recall that the interaction terms in the non-linear sigma model contain two derivatives.
Each diagram with the leading IR logarithms contains a closed loop of the twice differentiated propagators 
which runs through the vertex located at the external point $x$. 
The other diagrams are obtained if we remove any of the differential operators from the closed loop
and let them act on the other propagators outside the loop.
We can show that such diagrams always have reduced powers of the IR logarithms.
We explain the details of the IR power counting in non-linear sigma models in Appendix A.

Therefore in the "vertex" terms, the diagrams with the leading IR logarithms contain the following structure:  
\begin{align}
\text{(The "vertex" terms)}\simeq
\parbox{\boxxal}{\usebox{\boxxa}}
+\parbox{\boxxcl}{\usebox{\boxxc}}.
\label{nonlinear}\end{align}
To evaluate the leading IR effects of the "propagator" terms, 
we have only to consider the diagrams 
where $\partial_\rho\xi\partial_\sigma\xi$ is inserted to one of the propagators of such a loop: 
\begin{align}
\text{(The "propagator" terms)}\simeq
\parbox{\boxxbl}{\usebox{\boxxb}}
+\parbox{\boxxdl}{\usebox{\boxxd}}.
\label{linear}\end{align}
When the closed loop consists of a single propagator, we obtain 
\begin{align}
\parbox{\boxxbl}{\usebox{\boxxb}}
\simeq-\parbox{\boxxal}{\usebox{\boxxa}}. 
\label{ab}\end{align}
The important point is that there are equal number of the propagators and the vertices in a close loop.
The "vertex" terms count the vertices while the "propagator" terms count the propagators.
The "propagator" terms cancel the corresponding "vertex" terms.
To prove the cancellation in general, we focus on a pair of the corresponding terms: 
\begin{align}
\parbox{\boxxcl}{\usebox{\boxxc}}
=&\ F\int\sqrt{-g'}d^Dx'\ 
g^{\alpha\beta}(\tau')\sum_{i=\pm}sgn(+,i)\label{sNL}\\
&\times\cdots\partial_\varepsilon''''\partial_\rho G^{l+}(x'''',x)
\partial_\sigma\partial_\alpha' G^{+i}(x,x')
\partial_\beta'\partial_\zeta'''G^{ik}(x',x''')\cdots,\notag
\end{align}
\begin{align}
\parbox{\boxxdl}{\usebox{\boxxd}}
=&-iF\int\sqrt{-g''}d^Dx''\int\sqrt{-g'}d^Dx'\ 
g^{\gamma\delta}(\tau'')g^{\alpha\beta}(\tau')\sum_{i,j=\pm}sgn(j,+)sgn(+,i)\notag\\
&\hspace{-1em}\times\cdots\partial_\varepsilon''''\partial_\gamma''G^{lj}(x'''',x'')
\partial_\delta''\partial_\rho G^{j+}(x'',x)\partial_\sigma\partial_\alpha' G^{+i}(x,x')
\partial_\beta'\partial_\zeta'''G^{ik}(x',x''')\cdots,\label{sL}
\end{align}
where $F$ is a common coefficient between the "propagator" term and the "vertex " term
which is a function of covariant tensors such as $R_{cadb}$
and $sgn(i,j)$ is defined as
\begin{align}
sgn(i,j)\equiv
\begin{cases}+1&\text{for }(i,j)=(+,+),(-,-),\\-1&\text{for }(i,j)=(+,-),(-,+).\end{cases}
\end{align}
Note that (\ref{sL}) has the extra prefactor $-i$ compared with (\ref{sNL}). 
It is because the "propagator" terms have one more vertex than the "vertex" terms. 
By using the partial integration, (\ref{sL}) is
\begin{align}
\parbox{\boxxdl}{\usebox{\boxxd}}
\simeq&+iF\int d^Dx''\int\sqrt{-g'}d^Dx'\ 
g^{\alpha\beta}(\tau')\sum_{i,j=\pm}sgn(j,+)sgn(+,i)\label{sL1}\\
&\hspace{-1.5em}\times\cdots \partial_\varepsilon''''G^{lj}(x'''',x'')
\partial_\rho\sqrt{-g''}\nabla''^2 G^{j+}(x'',x)\partial_\sigma\partial_\alpha' G^{+i}(x,x')
\partial_\beta'\partial_\zeta'''G^{ik}(x',x''')\cdots, \notag
\end{align}
where we neglected the diagrams which don't induce the leading IR effects. 
By using (\ref{nabla}) and the partial integration, 
\begin{align}
\parbox{\boxxdl}{\usebox{\boxxd}}
\simeq&-F\int\sqrt{-g'}d^Dx'\ 
g^{\alpha\beta}(\tau')\sum_{i=\pm}sgn(+,i)\label{sL2}\\
&\times\cdots\partial_\varepsilon''''\partial_\rho G^{l+}(x'''',x)
\partial_\sigma\partial_\alpha' G^{+i}(x,x')
\partial_\beta'\partial_\zeta'''G^{ik}(x',x''')\cdots.\notag
\end{align}
Here we neglected the diagrams which don't induce the leading IR effects again. 
From (\ref{sNL}) and (\ref{sL2}), we obtain
\begin{align}
\parbox{\boxxdl}{\usebox{\boxxd}}
\simeq-\parbox{\boxxcl}{\usebox{\boxxc}}. 
\label{cd}\end{align}
This concludes the proof that the leading IR logarithms cancel in non-linear sigma models
to all orders.

\section{Sub-leading IR effects at the two loop level}
\setcounter{equation}{0}

In this section, we investigate the sub-leading IR effects to the cosmological constant at the two loop level. 

To perform the calculation efficiently, 
we note that the dS invariance is preserved up to the two loop level.  
It is because the leading IR effect: $\log a(\tau)$ is absent. 
So the vev of the energy-momentum tensor is written as
\begin{align}
\langle T_{\mu\nu}\rangle= \frac{g_{\mu\nu}}{D}\langle T_\rho^{\ \rho}\rangle. 
\end{align}
We have only to evaluate the trace of the energy-momentum tensor. 

In the non-linear sigma model, 
the trace of the energy-momentum tensor is 
\begin{align}
\langle T_\mu^{\ \mu}\rangle
=&\ (\frac{D}{2}-1)\langle-\{1+(\delta\beta+2\delta\gamma)R\}g^{\mu\nu}\partial_\mu\xi^a\partial_\nu\xi^a
+\frac{g^2}{3}R_{cadb}\xi^c\xi^dg^{\mu\nu}\partial_\mu\xi^a\partial_\nu\xi^b \rangle\label{NLT0}\\
=&\ (\frac{D}{2}-1)\langle-\{1+(\delta\beta+2\delta\gamma)R\}\frac{1}{2}\nabla^2(\xi^a\xi^a)
+\{1+(\delta\beta+2\delta\gamma)R\}\xi^a\nabla^2\xi^a\notag\\
&\hspace{4em}+\frac{g^2}{3}R_{cadb}\xi^c\xi^dg^{\mu\nu}\partial_\mu\xi^a\partial_\nu\xi^b \rangle\notag\\
=&\ (\frac{D}{2}-1)\langle-\{1+(\delta\beta+2\delta\gamma)R\}\frac{1}{2}\nabla^2(\xi^a\xi^a)\notag\\
&\hspace{4em}+\frac{g^2}{6}(R_{cadb}+R_{cbda})\xi^a\frac{1}{\sqrt{-g}}\partial_\mu(\xi^c\xi^d\sqrt{-g}g^{\mu\nu}\partial_\nu\xi^b) \rangle.\notag
\end{align}
In the third line of (\ref{NLT0}), we have used the equation of motion
\begin{align}
\{1+(\delta\beta+2\delta\gamma)R\}\nabla^2\xi^a
-\frac{g^2}{6}(R_{cadb}+R_{cbda})\frac{1}{\sqrt{-g}}\partial_\mu(\xi^c\xi^d\sqrt{-g}g^{\mu\nu}\partial_\nu\xi^b)&\\
+\frac{g^2}{6}(R_{cadb}+R_{cbda})\xi^bg^{\mu\nu}\partial_\mu\xi^c\partial_\nu\xi^d&=0. \notag
\end{align}
Up to the two loop level, 
\begin{align}
\langle T_\mu^{\ \mu}\rangle
=&-(\frac{D}{2}-1)\frac{1}{2}\{1+(\delta\beta+2\delta\gamma)R\}\nabla^2\langle\xi^a\xi^a\rangle\label{NLT}\\
&+(\frac{D}{2}-1)\frac{g^2}{6}(R_{cadb}+R_{cbda})\langle\xi^a \partial_\mu\xi^c\xi^dg^{\mu\nu}\partial_\nu\xi^b
+\xi^a\xi^c\partial_\mu\xi^dg^{\mu\nu}\partial_\nu\xi^b\rangle.\notag\\
=&-(\frac{D}{2}-1)\frac{1}{2}\nabla^2\langle\xi^a\xi^a\rangle
+(\frac{D}{2}-1)\frac{2g^2R}{3}\frac{H^{2D-2}}{(4\pi)^D}\frac{\Gamma^2(D-1)}{\Gamma^2(\frac{D}{2})}(D-1)\delta\notag\\
&+\frac{g^2RH^6}{2^5\pi^4}\log a(\tau)-\frac{g^2RH^6}{2^6\cdot 3\pi^4}.\notag
\end{align}
In the third line of (\ref{NLT}), we have used (\ref{counter}). 
To evaluate the sub-leading IR effects, we have to calculate the two point function up to $g^2\log a(\tau)$. 
In Appendix B, it is evaluated as:  
\begin{align}
\langle\xi^a\xi^a\rangle|_{g^2}\simeq&\ \frac{g^2RH^4}{2^5\cdot 3\pi^4}\big\{-\log^2a(\tau)+6(-2+\log 2+\gamma)\log a(\tau)\big\}\label{tworesult}\\
&+\frac{2g^2R}{3}\frac{H^{2D-4}}{(4\pi)^D}\frac{\Gamma^2(D-1)}{\Gamma^2(\frac{D}{2})}\delta\log a(\tau), \notag
\end{align}
where $\gamma$ is the Euler's constant. 
From (\ref{NLT}) and (\ref{tworesult}), the trace of the energy-momentum tensor up to the two loop level is
\begin{align}
\langle T_\mu^{\ \mu}\rangle=N\frac{3H^4}{8\pi^2}
&+(\frac{D}{2}-1)\frac{g^2RH^{2D-2}}{(4\pi)^D}\frac{\Gamma^2(D-1)}{\Gamma^2(\frac{D}{2})}(D-1)\delta\label{NLT1}\\
&-\frac{g^2RH^6}{2^6\pi^4}(13-6\log 2-6\gamma). \notag
\end{align}
At the two loop level, we have confirmed that the matter contribution to the cosmological constant is time independent. 
To obtain the time dependence of the effective cosmological constant, 
we have to investigate the sub-leading IR effects beyond the two loop level. 
In Section $7$, we investigate the sub-leading IR effects at the three loop level on an arbitrary target space. 
Before investigating it,  we consider the non-linear sigma model on an $S_N$ in the large $N$ limit in the next section. 

\section{Non-linear sigma model on $S_N$ in the large $N$ limit}
\setcounter{equation}{0}

In the case that the target space is an $S_N$, by introducing the auxiliary field $\chi$, 
the action of the non-linear sigma model is written as
\begin{align}
S_{matter}=\int\sqrt{-g}d^4x\ \big[-\frac{1}{2}g^{\mu\nu}\partial_\mu\varphi^i\partial_\nu\varphi^i
-\frac{\chi}{2}\big((\varphi^i)^2-\frac{1}{g^2}\big)\big], 
\end{align}
where $i=1\cdots N+1$. 
The field $\chi$ imposes the following constraint 
\begin{align}
(\varphi^i)^2=\frac{1}{g^2}. 
\label{constraint}\end{align}

In the large $N$ limit, we can neglect the fluctuation of $\chi$. 
So the action reduces to a free massive scalar field theory plus the constant term $\chi/g^2$. 
Here the auxiliary field is identified as the mass term: $\chi=m^2$. 

In order to satisfy the constraint (\ref{constraint}), 
we have to introduce the classical expectation value $(\varphi_{cl}^i(x))^2$ 
in addition to the quantum one $\langle(\tilde{\varphi}^i(x))^2\rangle$: 
\begin{align}
(\varphi^i)^2=(\varphi_{cl}^i(x))^2+\langle(\tilde{\varphi}^i(x))^2\rangle=\frac{1}{g^2}. 
\end{align}
It is because $1/g^2$ is a constant and
even if a scalar field is massive, its propagator is time dependent 
until $t\sim 3H/2m^2$ \cite{Vilenkin1982,Linde1982,Starobinsky1982}. 
At the coincident point, the propagator for a massive field is written as \cite{Miao2010}
\begin{align}
\langle(\tilde{\varphi}^i(x))^2\rangle
= &\ (N+1)\frac{H^{D-2}}{(4\pi)^\frac{D}{2}}\frac{\Gamma(1-\frac{D}{2})\Gamma(\frac{D-1}{2}+\nu)\Gamma(\frac{D-1}{2}-\nu)}{\Gamma(\frac{1}{2}+\nu)\Gamma(\frac{1}{2}-\nu)}\label{mpropagator}\\
&+(N+1)\frac{H^{D-2}}{(4\pi)^\frac{D}{2}}\frac{\Gamma(\nu)\Gamma(2\nu)}{\Gamma(\frac{D-1}{2})\Gamma(\frac{1}{2}+\nu)}\frac{(a(\tau))^{2\nu-(D-1)}}{\nu-\frac{D-1}{2}},\notag
\end{align}
where $\nu\equiv\sqrt{(D-1)^2/4-m^2/H^2}$ and we have adopted the assumption: $m^2/H^2\ll 1$. 
The UV divergence in (\ref{mpropagator}) is renormalizable  
by the coupling constant renormalization: $1/g^2\to 1/g^2-\delta g^2/g^4$. 

From (\ref{conservation}), the $g_{\mu\nu}$ term is always dominant in the energy-momentum tensor
irrespective of whether the dS invariance is respected or broken
\begin{align}
\langle T_{\mu\nu}\rangle\simeq\frac{g_{\mu\nu}}{D}\langle T_\rho^{\ \rho}\rangle. 
\end{align}
The trace of the energy-momentum tensor is
\begin{align}
\langle T_\mu^{\ \mu}\rangle
=&\langle-(\frac{D}{2}-1)g^{\mu\nu}\partial_\mu\varphi^i\partial_\nu\varphi^i
-\frac{D}{2}m^2\big((\varphi^i)^2-(\frac{1}{g^2}-\frac{\delta g^2}{g^4})\big)\rangle\label{NNLT}\\ 
=&\ (\frac{D}{2}-1)\langle-\frac{1}{2}\nabla^2(\varphi^i)^2+m^2(\varphi^i)^2\rangle\notag\\
=&\ (\frac{D}{2}-1)m^2(\frac{1}{g^2}-\frac{\delta g^2}{g^4}).\notag
\end{align}
Here we have used the constraint (\ref{constraint}) and the equation of motion
\begin{align}
\nabla^2\varphi^i-m^2\varphi^i=0. 
\end{align}

First, we confirm the result (\ref{NLT1}) in the leading order of $N$. 
To do so, we expand (\ref{mpropagator}) up to $\mathcal{O}(m^2/H^2)$
\begin{align}
\langle(\tilde{\varphi}^i(x))^2\rangle
=&\ (N+1)\frac{H^{D-2}}{(4\pi)^\frac{D}{2}}\frac{\Gamma(D-1)}{\Gamma(\frac{D}{2})}(2\log a(\tau)+\delta)\\
&+(N+1)\frac{m^2}{H^2}\Big[-\frac{H^2}{12\pi^2}\big\{\log^2a(\tau)+2(2-\log 2-\gamma)\log a(\tau)\big\}+X\Big]. \notag
\end{align}
Here $X$ denotes the UV divergent constant at $\mathcal{O}(m^2/H^2)$. 
To evaluate the two loop effect, we don't need to know its value. 
To renormalize the UV divergence at $t=0$, we choose the counter term as
\begin{align}
-\frac{\delta g^2}{g^4}=(N+1)\frac{H^{D-2}}{(4\pi)^\frac{D}{2}}\frac{\Gamma(D-1)}{\Gamma(\frac{D}{2})}\delta
+(N+1)\frac{m^2}{H^2}X,
\label{counterg}\end{align}
\begin{align}
(\varphi_{cl}^i(x))^2
=&\ \frac{1}{g^2}-(N+1)\frac{2H^{D-2}}{(4\pi)^\frac{D}{2}}\frac{\Gamma(D-1)}{\Gamma(\frac{D}{2})}\log a(\tau)\label{cl}\\
&+(N+1)\frac{m^2}{2^2\cdot 3\pi^2}\big\{\log^2a(\tau)+2(2-\log 2-\gamma)\log a(\tau)\big\}.\notag
\end{align}
By substituting (\ref{cl}) in the equation of motion
\begin{align}
\nabla^2\varphi^i_{cl}-m^2\varphi^i_{cl}=0,
\end{align}
we evaluate the mass term
\begin{align}
m^2=(N+1)g^2\frac{H^{D}}{(4\pi)^\frac{D}{2}}\frac{\Gamma(D)}{\Gamma(\frac{D}{2})}-\frac{(N+1)^2g^4H^6}{2^6\pi^4}(13-6\log 2-6\gamma). 
\label{m}\end{align}
The value at $\mathcal{O}(g^2)$ is consistent with the result in \cite{Davis1991}. 
Note that the assumption $m^2/H^2\ll 1$ is consistent if $Ng^2H^2\ll 1$. 
From (\ref{NNLT}), (\ref{counterg}) and (\ref{m}),  
\begin{align}
\langle T_\mu^{\ \mu}\rangle
=(N+1)\frac{3H^4}{8\pi^2}&+g^2(N+1)^2(\frac{D}{2}-1)\frac{H^{2D-2}}{(4\pi)^D}\frac{\Gamma^2(D-1)}{\Gamma^2(\frac{D}{2})}(D-1)\delta\\
&-\frac{g^2(N+1)^2H^6}{2^6\pi^4}(13-6\log 2-6\gamma).\notag 
\end{align}
As we recall $R=N(N-1)$ on an $S_N$, 
the result coincides with (\ref{NLT1}) in the leading order of $N$. 

Our interest is whether the effective cosmological constant becomes time dependent 
if we consider the higher loop effects. 
From (\ref{NNLT}) we find that the effective cosmological constant is time independent
as long as the effective mass is time independent. 
If the effective mass becomes time dependent, 
the energy-momentum tensor has the UV divergent term whose coefficient is time dependent. 
The counter terms are highly restricted in the non-linear sigma model on an $S_N$ in the large $N$ limit.
Since $\varphi^i\varphi^i$ is constrained to be a constant, possible scalar field dependent counter terms 
must contain $g^{\mu\nu}\partial_\mu\varphi^i\partial_\nu\varphi^i$.
In the large $N$ limit they must be bilinear in $\varphi^i$ with the indices $i$ contracted.
Time dependent UV-divergences cannot be
renormalized by the cosmological constant 
or possible other counter terms such as 
$R_g g^{\mu\nu}\partial_\mu\varphi^i\partial_\nu\varphi^i$ where $R_g$ is the scalar curvature
of dS space. The significance of this kind of counter term will be explained in the next section.
On the other hand, we expect the renormalizability to hold if we allow all possible counter terms.
Therefore we argue that  
the effective cosmological constant is time independent on an $S_N$ in the large $N$ limit
even if we consider the full IR effects.

\section{IR effects at the three loop level}
\setcounter{equation}{0}

Following the result in the previous section, 
it is natural to ask whether the effective cosmological constant has time dependence on a generic target space. 
As we have shown the cancellation of the leading IR logarithms to all orders, there is no $\log ^2a(\tau)$ type term at 
the three loop level. However there could still exist a sub-leading $\log a(\tau)$ type term 
in a generic non-linear sigma model.
In this section, we investigate such IR effects on a generic target space. 

From (\ref{NLEMT}), the vev of the energy-momentum tensor up to the three loop level is
\begin{align}
\langle T_{\mu\nu}\rangle=&\ (\delta_\mu^{\ \rho}\delta_\nu^{\ \sigma}-\frac{1}{2}g_{\mu\nu}g^{\rho\sigma})\times\\
&\langle \partial_\rho\xi^a\partial_\sigma\xi^a-\frac{g^2}{3}R_{cadb}\xi^c\xi^d\partial_\rho\xi^a\partial_\sigma\xi^b\notag\\
&+(-\frac{g^4}{20}D_eD_fR_{cadb}+\frac{2g^4}{45}R^g_{\ cad}R_{gebf})\xi^c\xi^d\xi^e\xi^f\partial_\rho\xi^a\partial_\sigma\xi^b\rangle. \notag
\end{align}
The contribution at the three loop level consists of the three kinds of diagrams 
\begin{align}
\langle T_{\mu\nu}\rangle=(\delta_\mu^{\ \rho}\delta_\nu^{\ \sigma}-\frac{1}{2}g_{\mu\nu}g^{\rho\sigma})\times
&\Big[(\text{The chain diagrams})+(\text{The circle diagrams})\label{3loopEMT}\\
&+(\text{The clover diagrams})\Big]. \notag
\end{align}
These diagrams are represented as
\begin{align}
(\text{The chain diagrams})&=-i\frac{g^4}{9}R^{ab}R_{ab}
\Big[\parbox{\chainal}{\usebox{\chaina}}+\parbox{\chainil}{\usebox{\chaini}}+\cdots\Big], \\
(\text{The circle diagrams})&=-i\frac{g^4}{6}R^{cadb}R_{cadb}
\Big[\parbox{\circleal}{\usebox{\circlea}}+\parbox{\circledl}{\usebox{\circled}}+\cdots\Big], \notag\\
(\text{The clover diagrams})&=(\frac{2g^4}{45}R^{ab}R_{ab}+\frac{g^4}{15}R^{cadb}R_{cadb}-\frac{g^4}{10}D^2R)
\Big[\parbox{\cloveral}{\usebox{\clovera}}+\parbox{\clovercl}{\usebox{\cloverc}}+\cdots\Big].  \notag
\end{align}
Unlike in Section 4, we explicitly factor out the coefficients which are combinations of
$R^{ab}R_{ab}$, $R^{cadb}R_{cadb}$, $D^2R$. 

First, we reconfirm the cancellation of the leading IR effects of $\mathcal{O}(\log^2 a(\tau))$. 
By using the partial integration, we find
\begin{align}
&\parbox{\chainal}{\usebox{\chaina}}+\parbox{\chainil}{\usebox{\chaini}}
=-2\parbox{\chaindl}{\usebox{\chaind}}-2\parbox{\chainll}{\usebox{\chainl}}
=\mathcal{O}(\log a(\tau)), \label{PI1}\\
&\parbox{\chainbl}{\usebox{\chainb}}+\parbox{\chainjl}{\usebox{\chainj}}
=-\parbox{\chainel}{\usebox{\chaine}}-\parbox{\chainfl}{\usebox{\chainf}}-\parbox{\chainml}{\usebox{\chainm}}-\parbox{\chainnl}{\usebox{\chainn}}
=\mathcal{O}(\log a(\tau)), \notag\\
&\parbox{\chaincl}{\usebox{\chainc}}+\parbox{\chainkl}{\usebox{\chaink}}
=-2\parbox{\chaingl}{\usebox{\chaing}}-2\parbox{\chainol}{\usebox{\chaino}}
=\mathcal{O}(\log a(\tau)), \notag\\
&\parbox{\chainhl}{\usebox{\chainh}}+\parbox{\chainpl}{\usebox{\chainp}}
=-2\parbox{\chaingl}{\usebox{\chaing}}-2\parbox{\chainql}{\usebox{\chainq}}
=\mathcal{O}(\log a(\tau)), \notag
\end{align}
\begin{align}
\parbox{\circleal}{\usebox{\circlea}}+\parbox{\circledl}{\usebox{\circled}}
=-2\parbox{\circlebl}{\usebox{\circleb}}-2\parbox{\circlefl}{\usebox{\circlef}}
=\mathcal{O}(\log a(\tau)), 
\label{PI2}\end{align}
\begin{align}
\parbox{\cloveral}{\usebox{\clovera}}+\parbox{\clovercl}{\usebox{\cloverc}}
=-4\parbox{\cloverbl}{\usebox{\cloverb}}-4\parbox{\cloverfl}{\usebox{\cloverf}}
=\mathcal{O}(\log a(\tau)). 
\label{PI3}\end{align}
From (\ref{PI1}), (\ref{PI2}) and (\ref{PI3}), we can show that the total of the diagrams in (\ref{3loopEMT}) 
doesn't have the leading IR effect. 
Note that the leading IR effects cancel pairwise between 
a "propagator" term and a "vertex" term in accord with our proof in Section 4.

Next, we investigate the sub-leading IR effect. 
In Section $6$, we have shown that the vev of the energy-momentum tensor has no time dependence on an $S_N$ 
in the large $N$ limit where
\begin{align}
R^{ab}R_{ab}=N(N-1)^2=\mathcal{O}(N^3),\hspace{1em}R^{cadb}R_{cadb}=2N(N-1)=\mathcal{O}(N^2),\hspace{1em}D^2R=0. 
\end{align}
Therefore, the result in the large $N$ limit implies the cancellation of the time dependence between 
the following diagrams
\begin{align}
-i\frac{g^4}{9}R^{ab}R_{ab}\Big[\parbox{\chainal}{\usebox{\chaina}}+\parbox{\chainil}{\usebox{\chaini}}+\cdots\Big]
+\frac{2g^4}{45}R^{ab}R_{ab}\Big[\parbox{\cloveral}{\usebox{\clovera}}+\parbox{\clovercl}{\usebox{\cloverc}}+\cdots\Big]
=\text{const}. 
\label{PIN}\end{align}
In order to investigate the sub-leading IR effect, we only need to consider the remaining diagrams. 
By using (\ref{PI2}) and (\ref{PI3}), the remaining diagrams are written as follows
\begin{align}
&-i\frac{g^4}{6}R^{cadb}R_{cadb}
\Big[-4\parbox{\circlebl}{\usebox{\circleb}}+\parbox{\circlecl}{\usebox{\circlec}}+\parbox{\circleel}{\usebox{\circlee}}-3\parbox{\circlefl}{\usebox{\circlef}}\label{3sub-leading}\\
&\hspace{8em}-\parbox{\circlegl}{\usebox{\circleg}}-2\parbox{\circlehl}{\usebox{\circleh}}+2\parbox{\circleil}{\usebox{\circlei}}\Big]\notag\\
&+(\frac{g^4}{15}R^{cadb}R_{cadb}-\frac{g^4}{10}D^2R)
\Big[-5\parbox{\cloverbl}{\usebox{\cloverb}}+2\parbox{\cloverdl}{\usebox{\cloverd}}
-\parbox{\cloverel}{\usebox{\clovere}}-6\parbox{\cloverfl}{\usebox{\cloverf}}\Big]. \notag
\end{align}

By using the partial integration, we find
\begin{align}
\parbox{\cloverfl}{\usebox{\cloverf}}=-\frac{1}{2}\parbox{\cloverdl}{\usebox{\cloverd}}-\parbox{\cloverel}{\usebox{\clovere}}. 
\label{PI4}\end{align}
From this identity, the clover diagrams of (\ref{3sub-leading}) are written as follows
\begin{align}
(\frac{g^4}{3}R^{cadb}R_{cadb}-\frac{g^4}{2}D^2R)
\Big[\parbox{\cloverdl}{\usebox{\cloverd}}-\parbox{\cloverbl}{\usebox{\cloverb}}+\parbox{\cloverel}{\usebox{\clovere}}\Big]. 
\label{clover1}\end{align}
The third diagram in the right hand side does not induce an IR logarithm:
\begin{align}
\parbox{\cloverel}{\usebox{\clovere}}=\text{const}. 
\label{const1}\end{align}
We can confirm its time independence without an detailed calculation
as explained in Appendix A. 
Thus the clover diagrams are estimated as
\begin{align}
(\frac{g^4}{3}R^{cadb}R_{cadb}-\frac{g^4}{2}D^2R)
\Big[\parbox{\cloverdl}{\usebox{\cloverd}}-\parbox{\cloverbl}{\usebox{\cloverb}}\Big]. 
\label{clover2}\end{align}

In a similar way, we investigate the circle diagrams of (\ref{3sub-leading}). 
By using the partial integration, we find
\begin{align}
\parbox{\circlefl}{\usebox{\circlef}}=-\parbox{\circlehl}{\usebox{\circleh}}-\parbox{\circleil}{\usebox{\circlei}}-i\parbox{\cloverfl}{\usebox{\cloverf}}. 
\label{PI5}\end{align}
From this identity, the circle diagrams are evaluated as
\begin{align}
&-i\frac{g^4}{6}R^{cadb}R_{cadb}
\Big[-4\parbox{\circlebl}{\usebox{\circleb}}+\parbox{\circlecl}{\usebox{\circlec}}+\parbox{\circleel}{\usebox{\circlee}}+3i\parbox{\cloverfl}{\usebox{\cloverf}}\label{circle1}\\
&\hspace{8em}-\parbox{\circlegl}{\usebox{\circleg}}+\parbox{\circlehl}{\usebox{\circleh}}+5\parbox{\circleil}{\usebox{\circlei}}\Big]. \notag
\end{align}
In addition, we find the following identities by using the partial integration
\begin{align}
\parbox{\circlebl}{\usebox{\circleb}}=-\frac{1}{2}\parbox{\circlecl}{\usebox{\circlec}}-i\frac{1}{2}\parbox{\cloverbl}{\usebox{\cloverb}}, 
\label{PI6}\end{align}
\begin{align}
\parbox{\circlehl}{\usebox{\circleh}}=-\parbox{\circleel}{\usebox{\circlee}}-\parbox{\circlegl}{\usebox{\circleg}}
-i\parbox{\cloverdl}{\usebox{\cloverd}}-i\parbox{\cloverel}{\usebox{\clovere}}-i\parbox{\cloverfl}{\usebox{\cloverf}}. 
\label{PI7}\end{align}
From the above relations and (\ref{PI4}), (\ref{circle1}) is
\begin{align}
&-i\frac{g^4}{6}R^{cadb}R_{cadb}
\Big[2i\parbox{\cloverbl}{\usebox{\cloverb}}-2i\parbox{\cloverdl}{\usebox{\cloverd}}-3i\parbox{\cloverel}{\usebox{\clovere}}\\
&\hspace{8em}+3\parbox{\circlecl}{\usebox{\circlec}}-2\parbox{\circlegl}{\usebox{\circleg}}+5\parbox{\circleil}{\usebox{\circlei}}\Big]. \notag
\end{align}
By using the power counting in Appendix A  like in (\ref{const1}), 
we can confirm the time independence of the following diagrams
\begin{align}
\parbox{\circlecl}{\usebox{\circlec}}=\parbox{\circlegl}{\usebox{\circleg}}=\parbox{\circleil}{\usebox{\circlei}}=\text{const}. 
\label{const2}\end{align}
So the circle diagrams are estimated as
\begin{align}
-\frac{g^4}{3}R^{cadb}R_{cadb}
\Big[\parbox{\cloverdl}{\usebox{\cloverd}}-\parbox{\cloverbl}{\usebox{\cloverb}}\Big]. 
\label{circle2}\end{align}
From (\ref{3sub-leading}), (\ref{clover2}) and (\ref{circle2}), we conclude that
the vev of the energy-momentum tensor at the three loop level is
\begin{align}
\langle T_{\mu\nu}\rangle|_{g^4}
\simeq(\delta_\mu^{\ \rho}\delta_\nu^{\ \sigma}-\frac{1}{2}g_{\mu\nu}g^{\rho\sigma})\times
-\frac{g^4}{2}D^2R
\Big[\parbox{\cloverdl}{\usebox{\cloverd}}-\parbox{\cloverbl}{\usebox{\cloverb}}\Big]. 
\label{clover3}\end{align}
Here $\simeq$ denotes the equality with respect to the time dependent terms. 
The sub-leading IR effects which are proportional to $R^{cadb}R_{cadb}$ cancel out each other. 
Unlike the leading IR effects, this cancellation takes place between the different kinds of diagrams, 
between the clover diagrams and the circle diagrams. 
On the other hand, only the clover diagrams have the coefficient $D^2R$. 
That is why the sub-leading IR logarithm is proportional to $D^2R$. 
Note that $D^2R$ vanishes on symmetric spaces such as an $S_N$. 
Therefore the time independence of the cosmological constant on an $S_N$ also holds with finite $N$ at the three loop level.  
Furthermore, we point out that the identity (\ref{PIN}) can be confirmed also by using the above diagramatic investigation.  

The contribution from the remaining two diagrams is evaluated in Appendix C as
\begin{align}
\langle T_{\mu\nu}\rangle|_{g^4}
\simeq\ g_{\mu\nu}g^4D^2R\frac{(D-1)(D-2)}{2}\frac{H^{3D-4}}{(4\pi)^\frac{3D}{2}}\frac{\Gamma^2(D-1)}{\Gamma(\frac{D}{2})}
\Big\{\frac{1}{\varepsilon}\log a(\tau)-\frac{7}{6}\log a(\tau)\Big\}. 
\label{DR}\end{align}
Note that the coefficient of $\log a(\tau)$ is UV divergent
and it is not renormalizable by the existing counter terms (\ref{beta}), (\ref{gamma}).  
The time dependent diagrams arising from (\ref{beta}) and (\ref{gamma}) are
\begin{align}
&i\frac{g^2}{3}(\delta\beta+2\delta\gamma)R^{ab}R_{ab}
\Big[2\parbox{\counteral}{\usebox{\countera}}
+\parbox{\boxcl}{\usebox{\boxc}}
+\parbox{\counterbl}{\usebox{\counterb}}
+\parbox{\countercl}{\usebox{\counterc}}
+\parbox{\counterdl}{\usebox{\counterd}}\Big]\\
&+g^2\delta\beta(\frac{1}{2}D^2R-\frac{1}{3}R^{ab}R_{ab})
\Big[\parbox{\boxcl}{\usebox{\boxc}}+\parbox{\boxal}{\usebox{\boxa}}\Big], \notag
\end{align}
where a small dot denotes the counter term insertion. 
By using the partial integration, we find
\begin{align}
\parbox{\counteral}{\usebox{\countera}}\simeq-\parbox{\boxcl}{\usebox{\boxc}},\hspace{1em}
\parbox{\boxcl}{\usebox{\boxc}}\simeq\parbox{\counterbl}{\usebox{\counterb}}\simeq-\parbox{\boxal}{\usebox{\boxa}},\hspace{1em}
\parbox{\countercl}{\usebox{\counterc}}\simeq-\parbox{\counterdl}{\usebox{\counterd}}. 
\end{align}
From these identities, the total contribution from (\ref{beta}) and (\ref{gamma}) is time independent
\begin{align}
\delta_{\beta,\gamma}\langle T_{\mu\nu}\rangle|_{g^4}\simeq 0. 
\end{align} 
It is why (\ref{DR}) is not renormalizable by (\ref{beta}) and (\ref{gamma}). 

This time dependent UV divergence can be renormalized by introducing the following counter term
\begin{align}
\delta_{\alpha} \mathcal{L}=\frac{\delta\alpha}{g^2}(R_g-D(D-1)H^2)
R_{ij}(\varphi)g^{\mu\nu}\partial_\mu\varphi^i\partial_\nu\varphi^j,
\label{counterR}\end{align}
where $R_g$ denotes the Ricci scalar of spacetime. 
The necessity of this kind of counter term in $\lambda \varphi^4$ theory has been pointed out in
\cite{Woodard2002}.
The only effect of the counter term is to modify the energy-momentum tensor as: 
\begin{align}
\delta_\alpha\langle T_{\mu\nu}\rangle
=-2\delta\alpha \big\{g_{\mu\nu}((D-1)H^2K+\nabla^2K)-\nabla_\mu\nabla_\nu K\big\}, 
\end{align}
\begin{align}
K=\langle R_{ab}g^{\mu\nu}\partial_\mu\xi^a\partial_\nu\xi^b
+(\frac{g^2}{2}D_cD_dR_{ab}-\frac{g^2}{3}R^{e}_{\ cad}R_{eb})\xi^c\xi^dg^{\mu\nu}\partial_\mu\xi^a\partial_\nu\xi^b\rangle. 
\label{K}\end{align}
In a similar way to the leading IR effect at the two loop level, 
we find that the following part of (\ref{K}) has no time dependence
\begin{align}
R_{ab}\langle g^{\mu\nu}\partial_\mu\xi^a\partial_\nu\xi^b\rangle|_{g^2}
-\frac{g^2}{3}R^{e}_{\ cad}R_{eb}\langle \xi^c\xi^dg^{\mu\nu}\partial_\mu\xi^a\partial_\nu\xi^b\rangle|_{g^0}\simeq 0. 
\label{consistent}\end{align}
We fix $\delta\alpha$ to renormalize the two loop matter contribution to the cosmological constant in (\ref{NLT1}): 
\begin{align}
-2\delta\alpha (D-1)H^2R_{ab}\langle g^{\mu\nu}\partial_\mu\xi^a\partial_\nu\xi^b\rangle|_{g^0}
=&-(\frac{D}{2}-1)\frac{g^2RH^{2D-2}}{(4\pi)^D}\frac{\Gamma^2(D-1)}{\Gamma^2(\frac{D}{2})}\frac{D-1}{D}\delta\\
&+\frac{g^2RH^6}{2^8\pi^4}(13-6\log 2-6\gamma)+\frac{g^2RH^6}{2^8\pi^4}C, \notag
\end{align}
where we have used 
$\nabla_\mu\langle g^{\rho\sigma}\partial_\rho\xi^a\partial_\sigma\xi^b\rangle|_{g^0}=0$. 
Note that there is a finite ambiguity $C$ when we renormalize the UV divergence.
In particular the two loop effect is completely canceled by the counter term up to $\mathcal{O}(\varepsilon^0)$ by setting $C=0$.  
The result is 
\begin{align}
\delta\alpha
=-\frac{D-2}{4D(D-1)}\frac{g^2H^{D-4}}{(4\pi)^\frac{D}{2}}\frac{\Gamma(D-1)}{\Gamma(\frac{D}{2})}\delta
+\frac{g^2}{2^6\cdot 3^2\pi^2}(13-6\log 2-6\gamma)+\frac{g^2}{2^6\cdot 3^2\pi^2}C. 
\end{align}
At the three loop level, this counter term gives rise to the the following time dependent term
\begin{align}
&-2\delta\alpha g_{\mu\nu}(D-1)H^2\times\frac{g^2}{2}D_cD_dR_{ab}\langle\xi^c\xi^dg^{\rho\sigma}\partial_\rho\xi^a\partial_\sigma\xi^b\rangle|_{g^0}\label{DRcounter}\\
\simeq&-g_{\mu\nu}g^4D^2R\frac{(D-2)(D-1)}{2D}\frac{H^{3D-4}}{(4\pi)^\frac{3D}{2}}\frac{\Gamma^3(D-1)}{\Gamma^3(\frac{D}{2})}\delta\log a(\tau)\notag\\
&+g_{\mu\nu}g^4D^2R\frac{H^8}{2^{11}\pi^6}(13-6\log 2-6\gamma)\log a(\tau)
+g_{\mu\nu}g^4D^2R\frac{CH^8}{2^{11}\pi^6}\log a(\tau), \notag
\end{align}
where we have used the fact that $\nabla_\mu\langle \xi^c\xi^dg^{\rho\sigma}\partial_\rho\xi^a\partial_\sigma\xi^b\rangle|_{g^0}$ is constant. 
From (\ref{DR}) and (\ref{DRcounter}), we find
\begin{align}
\langle T_{\mu\nu}^\text{total}\rangle|_{g^4}\simeq g_{\mu\nu}g^4D^2R\frac{CH^8}{2^{11}\pi^6}\log a(\tau). 
\end{align}
We have thus shown that the energy-momentum tensor can be renormalized up to the three loop level
with the counter terms we have identified.
The resultant time dependence of the cosmological constant is proportional to $D^2R$.
However it is also proportional to a finite subtraction ambiguity $C$.
Therefore there exists a renormalization scheme with $C=0$ in generic non-linear models which preserves the dS symmetry
up to the three loop level. 

\section{IR effects of a higher derivative interaction}
\setcounter{equation}{0}

In this paper we have shown there exists a cancellation mechanism among IR logarithms beyond the power counting
estimates in non-linear models on generic manifolds. The leading cancellation occurs between the "propagator" 
and "vertex" terms as there are one to one correspondences between them. This feature is specific to the
interaction terms with two derivatives. Therefore such a cancellation does not take place if we consider the
higher derivative interaction terms. In this section
we investigate a model with a higher derivative interaction term 
where the leading IR effects to the cosmological constant doesn't cancel out each other. 
We adopt the following model as a specific example: 
\begin{align}
S_{matter}=\int\sqrt{-g}d^4x\ 
\big[-\frac{1}{2}g^{\mu\nu}\partial_\mu\varphi^i\partial_\nu\varphi^i
-\frac{\lambda}{16N^2}(\varphi^i)^2(g^{\mu\nu}\partial_\mu\varphi^j\partial_\nu\varphi^j)^2\big], 
\end{align}
where $i=1\cdots N$. 
Note that we have also introduced the scalar field left intact by differential operators in the higher derivative
interaction term.  
In addition, we impose ${O}(N)$ symmetry on the action because it becomes exactly solvable in the large $N$ limit. 
The energy-momentum tensor is written as
\begin{align}
\langle T_{\mu\nu}\rangle=
&\ (\delta_\mu^{\ \rho}\delta_\nu^{\ \sigma}-\frac{1}{2}g_{\mu\nu}g^{\rho\sigma})
\langle\partial_\rho\varphi^i\partial_\sigma\varphi^i\rangle\label{NKT}\\
&+(\delta_\mu^{\ \rho}\delta_\nu^{\ \sigma}-\frac{1}{4}g_{\mu\nu}g^{\rho\sigma})
\langle\frac{\lambda}{4N^2}(\varphi^i)^2\partial_\rho\varphi^j\partial_\sigma\varphi^jg^{\alpha\beta}\partial_\alpha\varphi^k\partial_\beta\varphi^k\rangle.\notag
\end{align}
Note that the $g_{\mu\nu}$ dependences of the "propagator" term and the "vertex" term 
are different from those in the two derivative interaction models. 

The quantum corrections arise at the three loop level. 
The leading IR effects from the "vertex" term and the "propagator" term are
\begin{align}
&\frac{\lambda}{4N^2}\langle(\varphi^i)^2\partial_\rho\varphi^j\partial_\sigma\varphi^jg^{\alpha\beta}\partial_\alpha\varphi^k\partial_\beta\varphi^k\rangle|_{\lambda^0}
\label{NKNL}\\
\simeq&\ N\frac{\lambda}{4}G^{++}(x,x)
\lim_{x'\to x}\partial_\rho\partial_\sigma' G^{++}(x,x')g^{\alpha\beta}\partial_\alpha\partial_\beta' G^{++}(x,x')\notag\\
&+\frac{\lambda}{2}G^{++}(x,x)
\lim_{x'\to x}\partial_\rho\partial_\beta' G^{++}(x,x')g^{\alpha\beta}\partial_\alpha\partial_\sigma' G^{++}(x,x')\notag\\
\simeq&+g_{\rho\sigma}(N+\frac{1}{2})\frac{3^2\lambda H^{10}}{2^{12}\pi^6}\log a(\tau), \notag
\end{align}
\begin{align}
&\langle\partial_\rho\varphi^i\partial_\sigma\varphi^i\rangle|_{\lambda}\\
\simeq&-iN\frac{\lambda}{4}\int\sqrt{-g'}d^Dx'\ G^{++}(x',x')\lim_{x''\to x'}\partial_\alpha'\partial_\beta''G^{++}(x',x'')\notag\\
&\times g^{\alpha\beta}(\tau')g^{\gamma\delta}(\tau')
\big[\partial_\rho\partial_\gamma'G^{++}(x,x')\partial_\sigma\partial_\delta'G^{++}(x,x')
-\partial_\rho\partial_\gamma'G^{+-}(x,x')\partial_\sigma\partial_\delta'G^{+-}(x,x')\big]\notag\\
&-i\frac{\lambda}{2}\int\sqrt{-g'}d^Dx'\ G^{++}(x',x')\lim_{x''\to x'}\partial_\alpha'\partial_\delta''G^{++}(x',x'')\notag\\
&\times g^{\alpha\beta}(\tau')g^{\gamma\delta}(\tau')
\big[\partial_\rho\partial_\gamma'G^{++}(x,x')\partial_\sigma\partial_\beta'G^{++}(x,x')
-\partial_\rho\partial_\gamma'G^{+-}(x,x')\partial_\sigma\partial_\beta'G^{+-}(x,x')\big]. \notag
\end{align}
By using the partial integration and extracting the leading IR effects, the "propagator" term is
\begin{align}
&\langle\partial_\rho\varphi^i\partial_\sigma\varphi^i\rangle|_{\lambda}\\
\simeq&+iN\frac{\lambda}{4}\int d^Dx'\ G^{++}(x',x')\lim_{x''\to x'}\partial_\alpha'\partial_\beta''G^{++}(x',x'')\notag\\
&\times g^{\alpha\beta}(\tau')
\big[\partial_\rho G^{++}(x,x')\partial_\sigma\sqrt{-g'}\nabla'^2G^{++}(x,x')
-\partial_\rho G^{+-}(x,x')\partial_\sigma\sqrt{-g'}\nabla'^2G^{+-}(x,x')\big]\notag\\
&+i\frac{\lambda}{8}\int d^Dx'\ G^{++}(x',x')\lim_{x''\to x'}\partial_\alpha'\partial_\beta''G^{++}(x',x'')\notag\\
&\times g^{\alpha\beta}(\tau')
\big[\partial_\rho G^{++}(x,x')\partial_\sigma\sqrt{-g'}\nabla'^2G^{++}(x,x')
-\partial_\rho G^{+-}(x,x')\partial_\sigma\sqrt{-g'}\nabla'^2G^{+-}(x,x')\big]. \notag
\end{align}
Here we have used the fact : $\lim_{x''\to x'}\partial_\alpha'\partial_\beta''G^{++}(x',x'')=g_{\alpha\beta}(\tau')\times\text{const}$, and
\begin{align}
&\ g^{\alpha\delta}(\tau')g_{\alpha\beta}(\tau')g^{\gamma\delta}(\tau')\\
&\times\big[\partial_\rho\partial_\gamma'G^{++}(x,x')\partial_\sigma\partial_\beta'G^{++}(x,x')
-\partial_\rho\partial_\gamma'G^{+-}(x,x')\partial_\sigma\partial_\beta'G^{+-}(x,x')\big]\notag\\
=&\ \frac{1}{D}g^{\alpha\beta}(\tau')g_{\alpha\beta}(\tau')g^{\gamma\delta}(\tau')\notag\\
&\times\big[\partial_\rho\partial_\gamma'G^{++}(x,x')\partial_\sigma\partial_\delta'G^{++}(x,x')
-\partial_\rho\partial_\gamma'G^{+-}(x,x')\partial_\sigma\partial_\delta'G^{+-}(x,x')\big]. \notag
\end{align}
By using (\ref{nabla}) and the partial integration, 
\begin{align}
\langle\partial_\rho\varphi^i\partial_\sigma\varphi^i\rangle|_{\lambda}
\simeq&-(N+\frac{1}{2})\frac{\lambda}{4}G^{++}(x,x)
\lim_{x'\to x}\partial_\rho\partial_\sigma' G^{++}(x,x')g^{\alpha\beta}\partial_\alpha\partial_\beta' G^{++}(x,x')\label{NKL}\\
=&-g_{\rho\sigma}(N+\frac{1}{2})\frac{3^2\lambda H^{10}}{2^{12}\pi^6}\log a(\tau).\notag
\end{align}
By substituting (\ref{NKNL}) and (\ref{NKL}) in (\ref{NKT}), 
\begin{align}
\langle T_{\mu\nu}\rangle\simeq
g_{\mu\nu}N\frac{3H^4}{32\pi^2}+g_{\mu\nu}(N+\frac{1}{2})\frac{3^2\lambda H^{10}}{2^{12}\pi^6}\log a(\tau)
+a^2(\tau)\delta_\mu^{\ 0}\delta_\nu^{\ 0}(N+\frac{1}{2})\frac{3\lambda H^{10}}{2^{12}\pi^6}. 
\end{align}
Here we have evaluated the coefficient of the $\delta_\mu^{\ 0}\delta_\nu^{\ 0}$ term by the conservation law. 
Unlike the non-linear sigma model, 
the leading IR effect of the energy-momentum tensor is nonvanishing in this model. 
The effective cosmological constant decreases with cosmic evolution
\begin{align}
\Lambda_{eff}\simeq\Lambda
-\kappa N\frac{3H^4}{32\pi^2}-\kappa(N+\frac{1}{2})\frac{3^2\lambda H^{10}}{2^{12}\pi^6}\log a(\tau). 
\end{align}

The perturbation theory breaks down when $\lambda H^6 \log a(\tau )\sim 1$.
In such a situation we need to sum up all leading IR logarithms.
We can evaluate such a nonperturbative IR effect in the large $N$ limit. 
By using the auxiliary fields $\alpha, \beta$, the action is written as
\begin{align}
S_{matter}=\int\sqrt{-g}d^4x\ 
\big[-\frac{1}{2}(1+\alpha\beta)g^{\mu\nu}\partial_\mu\varphi^i\partial_\nu\varphi^i
-\frac{1}{2}\beta^2(\varphi^i)^2+N\sqrt{\frac{2}{\lambda}}\alpha\beta^2\big].  
\end{align}
By differentiating the action with respect to $\alpha, \beta$, 
\begin{align}
\alpha=\frac{1}{N}\sqrt{\frac{\lambda}{2}}(\varphi^i)^2,\hspace{1em}
\beta=\frac{1}{2N}\sqrt{\frac{\lambda}{2}}g^{\mu\nu}\partial_\mu\varphi^i\partial_\nu\varphi^i. 
\label{auxiliary}\end{align}

In the large $N$ limit, we can neglect the fluctuation of the auxiliary fields. 
So the action reduces to a free massive field theory plus the constant term $N\sqrt{2/\lambda}\alpha\beta^2$. 
We can evaluate the saturation value of the following vevs
\begin{align}
\langle(\varphi^i)^2\rangle\simeq&\ N\frac{3H^4}{8\pi^2\beta^2},\label{vevs}\\
\langle g^{\mu\nu}\partial_\mu\varphi^i\partial_\nu\varphi^i\rangle
=&\ \frac{1}{2}\nabla^2\langle(\varphi^i)^2\rangle-\langle\varphi^i\nabla^2\varphi^i\rangle\notag\\
=&-\frac{\beta^2}{1+\alpha\beta}\langle(\varphi^i)^2\rangle\notag\\
\simeq&\ \frac{-1}{1+\alpha\beta}N\frac{3H^4}{8\pi^2}. \notag
\end{align}
Here we have adopted the assumption: $\beta^2/H^2\ll 1$ and used the equation of motion
\begin{align}
(1+\alpha\beta)\nabla^2\varphi^i-\beta^2\varphi^i=0. 
\label{eom}\end{align}
From (\ref{vevs}), (\ref{auxiliary}) is written as
\begin{align}
\alpha\simeq\frac{1}{N}\sqrt{\frac{\lambda}{2}}\cdot N\frac{3H^4}{8\pi^2\beta^2},\hspace{1em}
\beta=\frac{1}{2N}\sqrt{\frac{\lambda}{2}}\cdot \frac{-1}{1+\alpha\beta}N\frac{3H^4}{8\pi^2}. 
\label{auxiliary1}\end{align}
By solving (\ref{auxiliary1}), 
\begin{align}
\alpha=\frac{4}{9}\sqrt{\frac{2}{\lambda}}\cdot \frac{8\pi^2}{3H^4},\hspace{1em}
\beta=-\frac{3}{2}\sqrt{\frac{\lambda}{2}}\cdot \frac{3H^4}{8\pi^2}. 
\label{auxiliary2}\end{align}
Furthermore, the trace of the energy-momentum tensor is written as
\begin{align}
\langle T_\mu^{\ \mu}\rangle
=&\ \langle-(1+\alpha\beta)g^{\mu\nu}\partial_\mu\varphi^i\partial_\nu\varphi^i-2\beta^2(\varphi^i)^2+4N\sqrt{\frac{2}{\lambda}}\alpha\beta^2\rangle\label{NKT1}\\
=&\ \langle-(1+\alpha\beta)\frac{1}{2}\nabla^2(\varphi^i)^2+(1+\alpha\beta)\varphi^i\nabla^2\varphi^i
-2\beta^2(\varphi^i)^2+4N\sqrt{\frac{2}{\lambda}}\alpha\beta^2\rangle\notag\\
=&\ \langle-\beta^2(\varphi^i)^2+4N\sqrt{\frac{2}{\lambda}}\alpha\beta^2\rangle. \notag
\end{align}
In the third line, we have used the equation of motion (\ref{eom}). 
From (\ref{vevs}), (\ref{auxiliary2}) and (\ref{NKT1}), 
\begin{align}
\langle T_\mu^{\ \mu}\rangle\simeq 3N\frac{3H^4}{8\pi^2}. 
\end{align}
The vev of the energy-momentum tensor is
\begin{align}
\langle T_{\mu\nu}\rangle\simeq g_{\mu\nu}N\frac{3H^4}{32\pi^2}+g_{\mu\nu}N\frac{3H^4}{16\pi^2}. 
\end{align}
Note that the difference from the free field value is not suppressed by the coupling constant. 
It is the result of the resummation of the leading IR logarithms to all orders. 
The effective cosmological constant decreases with cosmic evolution at the initial stage, 
while it is eventually saturated at the value
\begin{align}
\Lambda_{eff}=\Lambda-\kappa N\frac{3H^4}{32\pi^2}-\kappa N\frac{3H^4}{16\pi^2}. 
\end{align}

\section{Conclusion}
\setcounter{equation}{0}

In our previous paper \cite{Kitamoto2010}, we have investigated the non-linear sigma model in dS space. 
We have shown the leading infra-red effects to the cosmological constant cancel out each other 
at the two loop level on arbitrary target space. Furthermore they cancel to all orders in the large $N$ limit on an $S_N$. 
In this paper, we have extended these investigations.
We have shown the cancellation of the leading IR effects in non-linear sigma models 
on an arbitrary target space to all orders. 
In the large $N$ limit on an $S_N$, we have further shown that the effective cosmological constant is time independent 
even if we consider the full IR effects. 
Although sub-leading IR logarithm could arise at the three loop level in a generic non-linear sigma model,
we have shown that there is a renormalization scheme to cancel it. 

We may reflect these results as follows. 
As discussed in \cite{Polyakov}, 
the Schwinger-Keldysh formalism is necessary to evaluate the perturbative effects in de Sitter space. 
In this sense, our problem belongs to nonequilibrium physics. 
However  
it may be described by an Euclidean field theory on $S_4$
if an equilibrium state is eventually established. 
In fact, in the models with polynomial interactions, 
the eventual equilibrium state in the stochastic approach \cite{Starobinsky1994,Woodard2005} is recovered
by considering the zero mode dynamics in an Euclidean field theory \cite{Rajaraman2010}. 

If such a correspondence works in the non-linear sigma model, 
we may retain the zero mode in $G_{ij}(\varphi)$ 
and the nonzero modes in $g^{\mu\nu}\partial_\mu\varphi^i\partial_\nu\varphi^j$ 
to obtain the leading IR effects. 
In this approximation, the action is equal to the free field action
because $G_{ij}(\varphi)$ has no coordinate dependence and can be put to identity by rescaling the nonzero modes. 
This argument may explain why the leading IR effects to the cosmological constant cancel out each other. 
Furthermore, the action on an $S_N$ does not contain fields left intact by differential operators 
due to the constraint $(\varphi^i)^2=1/g^2$. 
So the effective cosmological constant is time independent because there is no contribution from the zero mode.  

The above nonperturbative considerations don't constrain the sub-leading IR effect on an arbitrary target space.  
We have investigated IR effects up to the three loop level where the sub-leading IR effect could induce time dependence. 
We have found that the sub-leading IR effect to the cosmological constant remains if $D^2R\not=0$ 
but its coefficient is UV divergent. We have identified a counter term which can cancel such a divergence.
Furthermore a natural counter term can cancel the IR logarithm completely. 
Therefore there is a renormalization scheme in a generic non-linear sigma model which preserves
dS symmetry up to the three loop level.

It should be noted that the above cancellations hold in the non-linear sigma model with two derivative interactions. 
In a general model with higher derivative interactions, 
the IR effects to the cosmological constant do not necessary cancel out each other. 
In fact, we have found that the cancellation of the leading IR effects does not take place 
in a field theory with higher derivative interactions.  
On dimensional grounds we expect a higher derivative interaction induces smaller effect than that
of the non-linear sigma model with two derivative interactions. In fact the power counting
argument is in accord with this intuition as the first IR logarithms are expected at the three and two loop
level respectively. The cancellation mechanism of the IR logarithms 
in non-linear sigma models changes this picture significantly.
In fact there is a renormalization scheme in non-linear sigma models which cancels IR logarithms up to the three
loop level. Surprisingly the higher derivative interaction produces larger IR effects contrary to the dimensional
estimates. They could eventually sum up to the quantity as large as the one loop effect just like in the large N limit.

To understand the eventual IR effects in the physical quantities, 
we have to evaluate the IR effects nonperturbatively. 
The large $N$ limit is available for some cases as is demonstrated in this paper. 
However we still don't know how to evaluate the nonperturbative IR effects in a general model with derivative interactions.  
Our results may be relevant to investigate possible dS symmetry breaking due to IR effects in quantum gravity.
It is because the gravitational field contain massless and minimally coupled modes \cite{Woodard1996}.  
When we consider the IR effects of gravity, 
an important question is to ask whether the IR effects emerge in the physical quantities or not
\cite{Tanaka2007,Tsamis2007,Urakawa2010,Higuchi2011,Miao2011,Giddings2010,Chialva2011}. 
The higher derivative interactions may play a nontrivial role in such a question as we find it the case in this paper.

\section*{Acknowledgments}

This work is supported in part by Grant-in-Aid for Scientific Research from the Ministry of Education, Science and Culture of Japan.
We would like to thank A. Higuchi, E. Komatsu, D. Lyth, T. Matsuda, S.P. Miao, T. Tanaka, S. Weinberg 
and R.P. Woodard for discussions and information. 

\appendix
\section{Power counting of $\log a(\tau)$}
\setcounter{equation}{0}

We can estimate the power of the IR logarithms induced by a diagram without a detailed calculation. 
Here we explain how to do it.  

First of all, we recall that the interaction vertices are located in the past light-cone of the
energy-momentum tensor. Since we are interested in logarithmically large contributions,
we can assume that the conformal time of the interaction vertices $\tau_i$ are hierarchically separated
$|\tau_1|\ll|\tau_2|\ll|\tau_3|\ll\cdots$.
In such a configuration the separations of the interaction vertices are almost always time-like
$|\tau_i -\tau_j| > |{\bf x}_i-{\bf x}_j|$.

For the power counting, 
we have only to focus on the following behavior of the constituents in the amplitude. 
The spacetime metric and the propagator at the coincident point 
show the following time dependence:
\begin{align}
g_{\alpha\beta}(\tau')\sim\frac{1}{\tau'^2},\hspace{1em}
\sqrt{-g'}g^{\alpha\beta}(\tau')\sim \frac{1}{\tau'^2},\hspace{1em}
G^{++}(x',x')\sim\log|\tau'|. 
\label{coincident2}\end{align} 
Concerning the retarded propagator $G^R(x,x')$ 
and  the symmetric propagators $\bar{G}(x,x')$ between the separated points,
we focus on the following behavior: 
\begin{align}
G^R(x,x')&\sim\theta(\tau-\tau')\theta\big((\tau-\tau')^2-|{\bf x}-{\bf x}'|^2\big)\label{separated2},\\
\bar{G}(x,x')&\sim\log \big((\tau-\tau')^2-|{\bf x}-{\bf x}'|^2\big).\notag
\end{align}
Note that they are functions of $\Delta x^2\equiv-(\tau-\tau')^2+|{\bf x}-{\bf x}'|^2$ except for the factor $\theta(\tau-\tau')$. 
The behavior of the differentiated propagators follow from (\ref{coincident2}) and (\ref{separated2})
except for the twice differentiated propagator at the coincident point: 
\begin{align}
\partial_\alpha' G^{++}(x',x')\sim\frac{1}{\tau'},
\label{coincident3}\end{align}
\begin{align}
\partial_\alpha G^R(x,x')&=-\partial_\alpha' G^R(x,x')\sim\theta(\tau-\tau')\partial_\alpha\theta(-\Delta x^2),\label{separated3}\\
\partial_\alpha\partial_\beta' G^R(x,x')&\sim\theta(\tau-\tau')\partial_\alpha\partial_\beta'\theta(-\Delta x^2),\notag\\
\partial_\alpha \bar{G}(x,x')&=-\partial_\alpha' \bar{G}(x,x')\sim\frac{1}{\Delta x},\notag\\
\partial_\alpha\partial_\beta' \bar{G}(x,x')&\sim\frac{1}{\Delta x^2}. \notag
\end{align}
We estimate the twice differentiated propagator at the coincident point as follows: 
\begin{align}
\lim_{x''\to x'}\partial_\alpha'\partial_\beta''G^{++}(x',x'')\sim\frac{1}{\tau'^2}. 
\label{coincident4}\end{align}
If we expand (\ref{separated2}) and (\ref{separated3}) in the power series of $|\bf{x}-\bf{x}'|/\tau-\tau'$ considering $\tau-\tau'>|\bf{x}-\bf{x}'|$, 
the spatial integration doesn't induce a logarithm. We thus obtain
\begin{align}
\int d^3x'\ G^R(x,x')&\sim\theta(\tau-\tau')\times(\tau-\tau')^3,\label{separated4}\\
\int d^3x'\ \partial_\alpha G^R(x,x')&=-\int d^3x'\ \partial_\alpha' G^R(x,x')\sim\theta(\tau-\tau')\times(\tau-\tau')^2,\notag\\
\int d^3x'\ \partial_\alpha\partial_\beta' G^R(x,x')&\sim\theta(\tau-\tau')\times(\tau-\tau'),\notag\\
\bar{G}(x,x')&\sim\log (\tau-\tau').\notag\\
\partial_\alpha \bar{G}(x,x')&=-\partial_\alpha' \bar{G}(x,x')\sim\frac{1}{\tau-\tau'},\notag\\
\partial_\alpha\partial_\beta' \bar{G}(x,x')&\sim\frac{1}{(\tau-\tau')^2}.\notag
\end{align}

In the above estimates, we have focued on the logarithm part of the propagator: 
\begin{align}
G(x,x')\sim\log (\Delta x^2). 
\end{align}
To be more precise, the propagator has the inverse square part in addition: 
\begin{align}
G(x,x')\sim\frac{\tau\tau'}{\Delta x^2}-\frac{1}{2}\log(\Delta x^2). 
\label{inverse}\end{align}
If we take the zeroth order of the expansion by $|\bf{x}-\bf{x}'|/\tau-\tau'$ and the differentiations with 
respect to time, 
the twice differentiated propagators have different asymptotic behavior with respect to
 $\tau$ and $\tau'$ in comparison with (\ref{separated4}): 
\begin{align}
\int d^3x'\ \partial_\alpha\partial_\beta' G^R(x,x')&\sim\theta(\tau-\tau')\times\frac{\tau\tau'}{\tau-\tau'}, \\
\partial_\alpha\partial_\beta' \bar{G}(x,x')&\sim\frac{\tau\tau'}{(\tau-\tau')^4}.\notag
\end{align}
It seems that the estimation (\ref{separated4}) is not entirely valid. 
Nevertheless it can be justified as we consider contributions from beyond the zeroth order
expansion by $|\bf{x}-\bf{x}'|/\tau-\tau'$.

As a concrete example, let us perform the power counting of the IR logarithms induced by the following two diagrams
\begin{align}
\parbox{\chainbl}{\usebox{\chainb}},\ 
\parbox{\circlegl}{\usebox{\circleg}}. 
\end{align}
The first diagram is written as
\begin{align}
\parbox{\chainbl}{\usebox{\chainb}}
\sim&\int\sqrt{-g'}g^{\alpha\beta}(\tau')d^4x'\ 
\lim_{x'\to x}\partial_\rho\partial_\sigma'G^{++}(x,x')\\
&\times\big[\partial_\alpha'G^R(x,x')\bar{G}(x,x')+\partial_\alpha'\bar{G}(x,x')G^R(x,x')
\big]\partial_\beta'G^{++}(x',x'). \notag
\end{align}
By using (\ref{coincident2}), (\ref{coincident3}), (\ref{coincident4}) and (\ref{separated4}), each integral is estimated as
\begin{align}
&\int\sqrt{-g'}g^{\alpha\beta}(\tau')d^4x'\ 
\lim_{x'\to x}\partial_\rho\partial_\sigma'G^{++}(x,x')
\partial_\alpha'G^R(x,x')\bar{G}(x,x')\partial_\beta'G^{++}(x',x')\label{2log1}\\
\sim&\ \frac{1}{\tau^2}\int^\tau\frac{d\tau'}{\tau'^2}\ \frac{(\tau-\tau')^2}{\tau'}\log(\tau-\tau')\notag\\
\sim&\ \frac{1}{\tau^2}\int^\tau\frac{d\tau'}{\tau'}\ \left\{\log|\tau'|\sum_{n=0}A_n\Big(\frac{\tau}{\tau'}\Big)^n+\sum_{n=1}B_n\Big(\frac{\tau}{\tau'}\Big)^n\right\}
\sim a^2(\tau)\log^2 a(\tau), \notag
\end{align}
\begin{align}
&\int\sqrt{-g'}g^{\alpha\beta}(\tau')d^4x'\ 
\lim_{x'\to x}\partial_\rho\partial_\sigma'G^{++}(x,x')
\partial_\alpha'\bar{G}(x,x')G^R(x,x')\partial_\beta'G^{++}(x',x')\label{2log2}\\
\sim&\ \frac{1}{\tau^2}\int^\tau\frac{d\tau'}{\tau'^2}\ \frac{(\tau-\tau')^2}{\tau'}\notag\\
\sim&\  \frac{1}{\tau^2}\int^\tau\frac{d\tau'}{\tau'}\sum_{n=0}C_n\Big(\frac{\tau}{\tau'}\Big)^n
\sim a^2(\tau)\log a(\tau). \notag
\end{align}
In the above expressions, we have expanded the integrands considering $|\tau |<|\tau' |$ where $A_n$, $B_n$, $C_n$ are constant coefficients. 
For the power counting of the IR logarithms, we have only to retain the zeroth order $n=0$.  
From (\ref{2log1}) and (\ref{2log2}), 
\begin{align}
\parbox{\chainbl}{\usebox{\chainb}}\sim a^2(\tau)\log^2 a(\tau).
\end{align}

The second diagram is written as
\begin{align}
\parbox{\circlegl}{\usebox{\circleg}}
\sim&\int\sqrt{-g'}g^{\alpha\beta}(\tau')d^4x'\sqrt{-g''}g^{\gamma\delta}(\tau'')d^4x''\\ 
&\times 2\partial_\rho G^R(x,x')\big[
\partial_\alpha'G^R(x',x'')\partial_\gamma''\bar{G}(x',x'')\partial_\beta'\partial_\delta''\bar{G}(x',x'')\notag\\
&\hspace{6em}+\partial_\alpha'\bar{G}(x',x'')\partial_\gamma''G^R(x',x'')\partial_\beta'\partial_\delta''\bar{G}(x',x'')\notag\\
&\hspace{6em}+\partial_\alpha'\bar{G}(x',x'')\partial_\gamma''\bar{G}(x',x'')\partial_\beta'\partial_\delta''G^R(x',x'')
\big]\partial_\sigma\bar{G}(x,x'')\notag\\
&+\int\sqrt{-g'}g^{\alpha\beta}(\tau')d^4x'\sqrt{-g''}g^{\gamma\delta}(\tau'')d^4x''\notag\\ 
&\times\partial_\rho G^R(x,x')
\partial_\alpha'\bar{G}(x',x'')\partial_\gamma''\bar{G}(x',x'')
\partial_\beta'\partial_\delta''\bar{G}(x',x'')
\partial_\sigma G^R(x,x''). \notag
\end{align}
By using (\ref{separated4}), each integral is estimated as
\begin{align}
&\int\sqrt{-g'}g^{\alpha\beta}(\tau')d^4x'\sqrt{-g''}g^{\gamma\delta}(\tau'')d^4x''\label{0log1}\\ 
&\times 2\partial_\rho G^R(x,x')\big[
\partial_\alpha'G^R(x',x'')\partial_\gamma''\bar{G}(x',x'')\partial_\beta'\partial_\delta''\bar{G}(x',x'')\notag\\
&\hspace{6em}+\partial_\alpha'\bar{G}(x',x'')\partial_\gamma''G^R(x',x'')\partial_\beta'\partial_\delta''\bar{G}(x',x'')\notag\\
&\hspace{6em}+\partial_\alpha'\bar{G}(x',x'')\partial_\gamma''\bar{G}(x',x'')\partial_\beta'\partial_\delta''G^R(x',x'')
\big]\partial_\sigma\bar{G}(x,x'')\notag\\
\sim&\int^\tau\frac{d\tau'}{\tau'^2}\int^{\tau'}\frac{d\tau''}{\tau''^2}\ 
(\tau-\tau')^2\frac{1}{(\tau'-\tau'')}\frac{1}{(\tau-\tau'')}\notag\\
\sim&\int^\tau d\tau'\int^{\tau'}\frac{d\tau''}{\tau''^4}\ 
\sum_{p,q,r=0}D_{pqr}\Big(\frac{\tau}{\tau'}\Big)^p\Big(\frac{\tau'}{\tau''}\Big)^q\Big(\frac{\tau}{\tau''}\Big)^r\sim a^2(\tau), \notag
\end{align}
\begin{align}
&\int\sqrt{-g'}g^{\alpha\beta}(\tau')d^4x'\sqrt{-g''}g^{\gamma\delta}(\tau'')d^4x''\label{0log2}\\ 
&\times\partial_\rho G^R(x,x')
\partial_\alpha'\bar{G}(x',x'')\partial_\gamma''\bar{G}(x',x'')
\partial_\beta'\partial_\delta''\bar{G}(x',x'')
\partial_\sigma G^R(x,x'')\notag\\
\sim&\int^\tau\frac{d\tau'}{\tau'^2}\int^{\tau'}\frac{d\tau''}{\tau''^2}\ 
(\tau-\tau')^2\frac{1}{(\tau'-\tau'')^4}(\tau-\tau'')^2\notag\\
\sim&\int^\tau d\tau'\int^{\tau'}\frac{d\tau''}{\tau''^4}\ 
\sum_{p,q,r=0}E_{pqr}\Big(\frac{\tau}{\tau'}\Big)^p\Big(\frac{\tau'}{\tau''}\Big)^q\Big(\frac{\tau}{\tau''}\Big)^r\sim a^2(\tau). \notag
\end{align}
In the second line of (\ref{0log2}), we have performed the integrals in the order $|\tau |<|\tau' |<|\tau'' |$. 
In the third line, we have expanded the integrands respecting this ordering where $D_{pqr}$, $E_{pqr}$ are constant coefficients. 
Just like the first diagram, we have only to retain the zeroth order $p=q=r=0$ for the power counting of the IR logarithms. 
From (\ref{0log1}) and (\ref{0log2}), we conclude the second diagram has no IR logarithms
\begin{align}
\parbox{\circlegl}{\usebox{\circleg}}\sim a^2(\tau). 
\end{align}

The power counting procedure of the IR logarithms is summarized as follows. 
In the first step, we estimate the  relevant behavior of the constituents of a diagram 
by using (\ref{coincident2}), (\ref{coincident3}), (\ref{coincident4}) and (\ref{separated4}). 
In the second step, we time order the integrations over the interaction points.
In the third step, we expand the integrand in the power series of the ratios of the conformal time
respecting the time ordering. 
For the power counting of the leading IR logarithms of a diagram, 
we have only to integrate the zeroth order of the expansion. 

In order to prove this statement, 
we first estimate the IR logarithms when there are no differential operators involved at the interaction point.
The integral over the location of an interaction vertex
induces some power of IR logarithms as:
\begin{align}
\int\sqrt{-g'}d^4x'\ A^R(x,x')\bar{B}(x',x'')
&\sim\log^{m-1}|\tau''|\log^{n+1}|\tau''|,
\label{counting1}\end{align}
where $|\tau |<| \tau''|<|\tau' |$. 
$A^R(x,x')$ consists of one retarded propagator $G^R(x,x')$ 
and $(m-1)$ symmetric propagators $\bar{G}(x,x')$. 
$\bar{B}(x,x')$ consists of $n$ symmetric propagators $\bar{G}(x,x')$.
We next estimate the effect of the minimal derivative coupling on the above estimate:
The $g^{\alpha\beta}(\tau')$ at the interaction vertex induces $\tau'^2$ behavior
and there are at least two derivatives involved. 
At the zeroth order, 
\begin{align}
g^{\alpha\beta}(\tau')(\partial,\partial',\partial'')^p\sim\frac{1}{\tau'^{p-2}}
,\hspace{1em}p\geq 2. 
\label{counting2}\end{align}
In the case $p=2$, the integral over time induces a single logarithm. 
However the differentiations on the symmetric propagators reduce the previous estimate of the power of the IR logarithms (\ref{counting1}). 
In the case $p>2$, the integral over time doesn't induce the IR logarithm 
and the power of the IR logarithms is less than (\ref{counting1}).
From (\ref{counting1}) and (\ref{counting2}), 
we find that each integral doesn't induce the positive power of the scale setting conformal time $\tau''$. 
For the power counting of the leading IR logarithms of a diagram, 
we have only to integrate the leading order of the expansion. 
We can iteratively continue this argument to cover the whole amplitude.

Finally we prove that 
each diagram with the leading IR logarithms contains a closed loop of the twice differentiated propagators 
which runs through the vertex located at the external point $x$.
Each vertex integral of the closed loop corresponds to the $p=4$ case in (\ref{counting2}). 
If the closed loop has $n$ vertices, 
the leading IR logarithms comes from 
the case that the closed loop has $n$ retarded propagators and one symmetric propagator: 
\begin{align}
&\int\sqrt{-g(\tau_1)}g^{\alpha_1\beta_1}(\tau_1)d^4x_1\cdots
\int\sqrt{-g(\tau_n)}g^{\alpha_n\beta_n}(\tau_n)d^4x_n\label{closed}\\
&\times\partial_{\rho}\partial_{\alpha_1}G^R(x,x_1)\cdots
\partial_{\beta_{n-1}}\partial_{\alpha_n}G^R(x_{n-1},x_n)
\partial_{\beta_n}\partial_{\sigma}\bar{G}(x_n,x)\times L(x,x_1,\cdots,x_n), \notag
\end{align}
where $L(x,x_1,\cdots,x_n)$ is some powers of the IR logarithm induced outside the closed loop. 
To be exact, the closed loop contains other permutations of propagators. 
The investigation of them can be performed in a similar way. 
We have only to estimate the integrand of (\ref{closed}) at the zeroth order
\begin{align}
\int\sqrt{-g(\tau_1)}g^{\alpha_1\beta_1}(\tau_1)d^4x_1\cdots
\int\sqrt{-g(\tau_n)}g^{\alpha_n\beta_n}(\tau_n)d^4x_n
\sim\int\frac{d\tau_1}{\tau_1^2}\cdots\int\frac{d\tau_n}{\tau_n^2},
\end{align}
\begin{align}
&\partial_{\rho}\partial_{\alpha_1}G^R(x,x_1)\cdots
\partial_{\beta_{n-1}}\partial_{\alpha_n}G^R(x_{n-1},x_n)
\partial_{\beta_n}\partial_{\sigma}\bar{G}(x_n,x)\\
\sim&\ \theta(\tau-\tau_1)\cdots\theta(\tau_{n-1}-\tau_n)\times\tau_1\cdots\tau_n\times\frac{1}{\tau_n^2}, \notag
\end{align}
\begin{align}
L(x,x_1,\cdots,x_n)\sim\log^{p}|\tau|\log^{p_1}|\tau_1|\cdots\log^{p_n}|\tau_n|, 
\end{align}
where $p,p_1,\cdots,p_n\geq 0$. 
So the integration (\ref{closed}) is estimated as
\begin{align}
&\int\sqrt{-g(\tau_1)}g^{\alpha_1\beta_1}(\tau_1)d^4x_1\cdots
\int\sqrt{-g(\tau_n)}g^{\alpha_n\beta_n}(\tau_n)d^4x_n\label{leading}\\
&\times\partial_{\rho}\partial_{\alpha_1}G^R(x,x_1)\cdots
\partial_{\beta_{n-1}}\partial_{\alpha_n}G^R(x_{n-1},x_n)
\partial_{\beta_n}\partial_{\sigma}\bar{G}(x_n,x)\times L(x,x_1,\cdots,x_n)\notag\\
\sim&\ a^2(\tau)\big(\log a(\tau)\big)^{p+p_1+\cdots+p_n}. \notag
\end{align}
Here the IR logarithms are induced by $L(x,x_1,\cdots,x_n)$ 
and the closed loop doesn't induce the IR logarithms. 

The other diagrams are obtained if we remove any of the differential operators from the closed loop. 
As an example, we consider the diagram with the closed loop where one differential operator is removed. 
Such a differential operator acts on the IR logarithms outside the closed loop $L(x,x_1,\cdots,x_n)$. 
On the other hand, the closed loop doesn't induce the IR logarithms. 
Therefore the power of this diagram is one less than (\ref{leading}): 
\begin{align}
&\int\sqrt{-g(\tau_1)}g^{\alpha_1\beta_1}(\tau_1)d^4x_1\cdots
\int\sqrt{-g(\tau_n)}g^{\alpha_n\beta_n}(\tau_n)d^4x_n\label{sub-leading}\\
&\times\partial_{\rho}G^R(x,x_1)\cdots
\partial_{\beta_{n-1}}\partial_{\alpha_n}G^R(x_{n-1},x_n)
\partial_{\beta_n}\partial_{\sigma}\bar{G}(x_n,x)\times \partial_{\alpha_1}L(x,x_1,\cdots,x_n)\notag\\
\sim&\ a^2(\tau)\big(\log a(\tau)\big)^{p+p_1+\cdots+p_n-1}. \notag
\end{align}
In the case where we remove any other differential operator from the closed loop,
the power of the IR logarithms induced by the corresponding diagram is also one less than (\ref{leading}). 

If we remove two differential operators from the closed loop 
of the twice differentiated propagators, 
it is possible that the closed loop induces a single IR logarithm more than otherwise. 
However in this case, the part outside the closed loop induces two less power of the IR logarithm. 
Therefore, also in this case, 
the power of the IR logarithm is one less than (\ref{leading}).

Even if we remove more than two differential operators from the closed loop, we can similarly conclude that
the power of the IR logarithm induced by the corresponding diagram is less than (\ref{leading}). 

\section{Two point function at the two loop level}
\setcounter{equation}{0}

Here we explain how to calculate the two point function up to $g^2\log a(\tau)$. 
In this Appendix and the next, 
we apply the procedure developed in \cite{Woodard2002}.  
The two point function at the two loop level is written as
\begin{align}
\langle\xi^a\xi^a\rangle|_{g^2}
=&\ \int \sqrt{-g'}d^Dx'\ 
\ i\frac{g^2}{3}R\lim_{x''\to x'} \partial_\alpha'\partial_\beta'' G^{++}(x',x'')\\
&\times g^{\alpha\beta}(\tau')
\big[G^{++}(x,x')G^{++}(x,x')-G^{+-}(x,x')G^{+-}(x,x')\big]\notag\\
&-\int \sqrt{-g'}d^Dx'\ 
\ i\frac{g^2}{6}R\partial_\alpha' G^{++}(x',x')\notag\\
&\times g^{\alpha\beta}(\tau')
\partial_\beta' \big[G^{++}(x,x')G^{++}(x,x')-G^{+-}(x,x')G^{+-}(x,x')\big]\notag\\
&+\int \sqrt{-g'}d^Dx'\ 
\big\{i\frac{g^2}{3}RG^{++}(x',x')-i(\delta\beta+2\delta\gamma)R\big\}\notag\\
&\times g^{\alpha\beta}(\tau')
\big[\partial_\alpha' G^{++}(x,x')\partial_\beta' G^{++}(x,x')
-\partial_\alpha' G^{+-}(x,x')\partial_\beta' G^{+-}(x,x')\big]. \notag
\end{align}
By using the partial integration, 
\begin{align}
\langle\xi^a\xi^a\rangle|_{g^2}
=&\ i\frac{g^2}{3}R\int \sqrt{-g'}d^Dx'
\ \lim_{x''\to x'} \partial_\alpha'\partial_\beta'' G^{++}(x',x'')\label{two}\\
&\times g^{\alpha\beta}(\tau')
\big[G^{++}(x,x')G^{++}(x,x')-G^{+-}(x,x')G^{+-}(x,x')\big]\notag\\
&-i\frac{g^2}{3}R\int \sqrt{-g'}d^Dx'
\ \partial_\alpha' G^{++}(x',x')\notag\\
&\times g^{\alpha\beta}(\tau')
\partial_\beta' \big[G^{++}(x,x')G^{++}(x,x')-G^{+-}(x,x')G^{+-}(x,x')\big]\notag\\
&+\frac{g^2RH^4}{2^4\cdot 3\pi^4}\log^2a(\tau)
+\frac{2g^2R}{3}\frac{H^{2D-4}}{(4\pi)^D}\frac{\Gamma^2(D-1)}{\Gamma^2(\frac{D}{2})}\delta\log a(\tau). \notag
\end{align}
To evaluate the two point function, 
we have to calculate the remaining integrals up to $g^2\log a(\tau)$. 
From (\ref{G})-(\ref{Gdetail}), we evaluate the propagators at the coincident point in $D$ dimension 
\begin{align}
&\lim_{x''\to x'} \partial_\alpha'\partial_\beta'' G^{++}(x',x'')=-\frac{H^D}{2(4\pi)^\frac{D}{2}}\frac{\Gamma(D)}{\Gamma(\frac{D}{2}+1)}g_{\alpha\beta}(\tau'),\label{coincident}\\
&\partial_\alpha' G^{++}(x',x')=\frac{2H^{D-1}}{(4\pi)^\frac{D}{2}} \frac{\Gamma(D-1)}{\Gamma(\frac{D}{2})}a(\tau')\delta_\alpha^{\ 0}. \notag
\end{align}
To evaluate the integrals up to $g^2\log a(\tau)$, 
we may set $D=4$ and extract the following part from the propagator at the separated point
\begin{align}
G(x,x')G(x,x')\simeq\Big(\frac{H^2}{4\pi^2}\Big)^2 
(-\frac{1}{y}\log H^2\Delta x^2+\frac{1}{4}\log^2H^2\Delta x^2-(1-\gamma)\log H^2\Delta x^2). 
\label{separated}\end{align}
From (\ref{coincident}) and (\ref{separated}), the integrals are
\begin{align}
&\ i\frac{g^2}{3}R\int \sqrt{-g'}d^Dx'
\ \lim_{x''\to x'} \partial_\alpha'\partial_\beta'' G^{++}(x',x'')\label{two1}\\
&\times g^{\alpha\beta}(\tau')
\big[G^{++}(x,x')G^{++}(x,x')-G^{+-}(x,x')G^{+-}(x,x')\big]\notag\\
=&-i\frac{g^2RH^8}{2^7\pi^6}\int d^4x'\ a^4(\tau')\notag\\
&\times\big[(-\frac{1}{y_{++}}\log H^2\Delta x_{++}^2+\frac{1}{4}\log^2H^2\Delta x_{++}^2-(1-\gamma)\log H^2\Delta x_{++}^2)\notag\\
&\hspace{1em}-(-\frac{1}{y_{+-}}\log H^2\Delta x_{+-}^2+\frac{1}{4}\log^2H^2\Delta x_{+-}^2-(1-\gamma)\log H^2\Delta x_{+-}^2)\big], \notag
\end{align}
\begin{align}
&-i\frac{g^2}{3}R\int \sqrt{-g'}d^Dx'
\ \partial_\alpha' G^{++}(x',x')\label{two2}\\
&\times g^{\alpha\beta}(\tau')
\partial_\beta' \big[G^{++}(x,x')G^{++}(x,x')-G^{+-}(x,x')G^{+-}(x,x')\big]\notag\\
=&+i\frac{g^2RH^7}{2^6\cdot 3\pi^6}\int d^4x'\ a^3(\tau')\notag\\
&\times\partial_0'\big[(-\frac{1}{y_{++}}\log H^2\Delta x_{++}^2+\frac{1}{4}\log^2H^2\Delta x_{++}^2-(1-\gamma)\log H^2\Delta x_{++}^2)\notag\\
&\hspace{2em}-(-\frac{1}{y_{+-}}\log H^2\Delta x_{+-}^2+\frac{1}{4}\log^2H^2\Delta x_{+-}^2-(1-\gamma)\log H^2\Delta x_{+-}^2)\big], \notag
\end{align}
where $\Delta x^2_{++},\Delta x^2_{+-}$ and $y_{++},y_{+-}$ are
\begin{align}
&\Delta x^2_{++}=-(|\tau-\tau'|-ie)^2+({\bf x}-{\bf x}')^2,\\
&\Delta x^2_{+-}=-(\tau-\tau'+ie)^2+({\bf x}-{\bf x}')^2, \notag
\end{align}
\begin{align}
&y_{++}=H^2a(\tau)a(\tau')\Delta x^2_{++},\\
&y_{+-}=H^2a(\tau)a(\tau')\Delta x^2_{+-}. \notag
\end{align}
For example, we calculate the following integral
\begin{align}
&\int d^4x'\ a^4(\tau')\big[-\frac{1}{y_{++}}\log H^2\Delta x_{++}^2+\frac{1}{y_{+-}}\log H^2\Delta x_{+-}^2\big]\label{ex}\\
=&-\frac{a^{-1}(\tau)}{H^2}\int d^4x'\ a^3(\tau')\big[\frac{1}{\Delta x_{++}^2}\log H^2\Delta x_{++}^2-\frac{1}{\Delta x_{+-}^2}\log H^2\Delta x_{+-}^2\big]. \notag
\end{align}
The integrand can be represented as the derivative of a polynomial in logarithms
\begin{align}
\frac{1}{\Delta x^2}\log H^2\Delta x^2=\frac{1}{8}\partial^2(\log^2H^2\Delta x^2-2\log H^2\Delta x^2), 
\label{log1}\end{align}
and the logarithm is divided by the real part and the imaginary part
\begin{align}
\log(H^2\Delta x^2_{++})&=\log(H^2|\Delta \tau^2-r^2|)+i\pi\theta(\Delta \tau^2-r^2), \label{log2}\\
\log(H^2\Delta x^2_{+-})&=\log(H^2|\Delta \tau^2-r^2|)-i\pi\theta(\Delta \tau^2-r^2)\{\theta(\Delta\tau)-\theta(-\Delta\tau)\}, \notag
\end{align}
where $\Delta\tau\equiv\tau-\tau'$ and $r^2\equiv({\bf x}-{\bf x}')^2$. 
By substituting (\ref{log1}) and (\ref{log2}) in (\ref{ex}), 
\begin{align}
&\int d^4x'\ a^4(\tau')\big[-\frac{1}{y_{++}}\log H^2\Delta x_{++}^2+\frac{1}{y_{+-}}\log H^2\Delta x_{+-}^2\big]\label{ex1}\\
=&\ \frac{a^{-1}(\tau)}{H^2}\cdot 2i\pi^2\partial_0^2\int^\tau_{-\frac{1}{H}}d\tau'\ a^3(\tau')\int^{\Delta \tau}_0r^2dr\ \big\{\log H^2(\Delta \tau^2-r^2)-1\big\}\notag\\
=&\ \frac{a^{-1}(\tau)}{H^2}\cdot 2i\pi^2\partial_0^2\int^\tau_{-\frac{1}{H}}d\tau'\ a^3(\tau')\Delta\tau^3\big\{\frac{2}{3}\log 2H\Delta\tau-\frac{11}{9}\big\}\notag\\
=&\ \frac{a^{-1}(\tau)}{H^4}\cdot 8i\pi^2\int^{a(\tau)}_1da(\tau')\ (1-\frac{a(\tau')}{a(\tau)})\big\{-\log\frac{a(\tau')}{2}-\sum_{n=1}\frac{1}{n}\frac{a^n(\tau')}{a^n(\tau)}-1\big\}\notag\\
\simeq&-\frac{4i\pi^2}{H^4}\log a(\tau).  \notag
\end{align}
In a similar way, 
\begin{align}
&\int d^4x'\ a^4(\tau')\big[\frac{1}{4}\log^2 H^2\Delta x_{++}^2-\frac{1}{4}\log H^2\Delta x_{+-}^2\big]\label{ex2}\\
\simeq&\ \frac{4i\pi^2}{3H^4}\big\{-\log^2 a(\tau)+(2\log 2+1)\log a(\tau)\big\}, \notag
\end{align}
\begin{align}
&\int d^4x'\ a^4(\tau')\big[-(1-\gamma)\log H^2\Delta x_{++}^2+(1-\gamma)\log H^2\Delta x_{+-}^2\big]\label{ex3}\\
\simeq&-(1-\gamma)\frac{8i\pi^2}{3H^4}\log a(\tau), \notag
\end{align}
\begin{align}
&\int d^4x'\ a^3(\tau')\partial_0'\big[-\frac{1}{y_{++}}\log H^2\Delta x_{++}^2+\frac{1}{y_{+-}}\log H^2\Delta x_{+-}^2\big]\label{ex4}\\
\simeq&\ \frac{12i\pi^2}{H^3}\log a(\tau), \notag
\end{align}
\begin{align}
&\int d^4x'\ a^3(\tau')\partial_0'\big[\frac{1}{4}\log^2 H^2\Delta x_{++}^2-\frac{1}{4}\log H^2\Delta x_{+-}^2\big]\label{ex5}\\
\simeq&\ \frac{4i\pi^2}{H^3}\big\{\log^2 a(\tau)-(2\log 2+1)\log a(\tau)\big\}, \notag
\end{align}
\begin{align}
&\int d^4x'\ a^3(\tau')\partial_0'\big[-(1-\gamma)\log H^2\Delta x_{++}^2+(1-\gamma)\log H^2\Delta x_{+-}^2\big]\label{ex6}\\
\simeq&\ (1-\gamma)\frac{8i\pi^2}{H^3}\log a(\tau). \notag
\end{align}
From (\ref{two1}), (\ref{two2}) and (\ref{ex1})-(\ref{ex6}), 
\begin{align}
&\ i\frac{g^2}{3}R\int \sqrt{-g'}d^Dx'
\ \lim_{x''\to x'} \partial_\alpha'\partial_\beta'' G^{++}(x',x'')\label{two11}\\
&\times g^{\alpha\beta}(\tau')
\big[G^{++}(x,x')G^{++}(x,x')-G^{+-}(x,x')G^{+-}(x,x')\big]\notag\\
\simeq&\ \frac{g^2RH^4}{2^5\cdot 3\pi^4}\big\{-\log^2a(\tau)+2(\log 2-2+\gamma)\log a(\tau)\big\},\notag
\end{align}
\begin{align}
&-i\frac{g^2}{3}R\int \sqrt{-g'}d^Dx'
\ \partial_\alpha' G^{++}(x',x')\label{two22}\\
&\times g^{\alpha\beta}(\tau')
\partial_\beta' \big[G^{++}(x,x')G^{++}(x,x')-G^{+-}(x,x')G^{+-}(x,x')\big]\notag\\
\simeq&\ \frac{g^2RH^4}{2^4\cdot 3\pi^4}\big\{-\log^2a(\tau)+2(\log 2-2+\gamma)\log a(\tau)\big\}.\notag
\end{align}
By substituting (\ref{two11}) and (\ref{two22}) in (\ref{two}), 
\begin{align}
\langle\xi^a\xi^a\rangle|_{g^2}\simeq&\ \frac{g^2RH^4}{2^5\cdot 3\pi^4}\big\{-\log^2a(\tau)+6(-2+\log 2+\gamma)\log a(\tau)\big\}\\
&+\frac{2g^2R}{3}\frac{H^{2D-4}}{(4\pi)^D}\frac{\Gamma^2(D-1)}{\Gamma^2(\frac{D}{2})}\delta\log a(\tau). \notag
\end{align}

\section{Derivation of (\ref{DR})}
\setcounter{equation}{0}

Here we explain how to derive (\ref{DR}). 
From (\ref{G})-(\ref{Gdetail}), the second diagram in (\ref{clover3}) can be easily evaluated
\begin{align}
-\parbox{\cloverbl}{\usebox{\cloverb}}
=-\frac{1}{4}G^{++}(x,x)\partial_\rho G^{++}(x,x)\partial_\sigma G^{++}(x,x)
\simeq -a^2(\tau)\delta_\rho^{\ 0}\delta_\sigma^{\ 0}\frac{H^8}{2^8\pi^6}\log a(\tau). 
\label{int0}\end{align}

The contribution from the first diagram is written as
\begin{align}
\parbox{\cloverdl}{\usebox{\cloverd}}
=&-i\int\sqrt{-g'}d^Dx'\ G^{++}(x',x')g^{\alpha\beta}(\tau')\lim_{x''\to x'}\partial_\alpha'\partial_\beta''G^{++}(x',x'')\label{int1}\\
&\times\big[\partial_\rho G^{++}(x,x')\partial_\sigma G^{++}(x,x')-\partial_\rho G^{+-}(x,x')\partial_\sigma G^{+-}(x,x')\big]. \notag
\end{align}
From (\ref{G})-(\ref{Gdetail}), we find
\begin{align}
&G^{++}(x',x')=\frac{H^{D-2}}{(4\pi)^\frac{D}{2}}\frac{\Gamma(D-1)}{\Gamma(\frac{D}{2})}(2\log a(\tau')+\delta), \label{G1}\\
&g^{\alpha\beta}(\tau')\lim_{x''\to x'}\partial_\alpha'\partial_\beta''G^{++}(x',x'')=-\frac{H^D}{(4\pi)^\frac{D}{2}}\frac{\Gamma(D)}{\Gamma(\frac{D}{2})}, \notag
\end{align}
\begin{align}
\partial_\rho G(x,x')\partial_\sigma G(x,x')
=&\frac{4^{D-2}H^{2D-2}}{(4\pi)^D}\Gamma^2(\frac{D}{2})a^2(\tau)\label{G2}\\
&\times\big[4H^2a^2(\tau')\frac{\Delta x_\rho\Delta x_\sigma}{y^{4-\varepsilon}}
+(4-\varepsilon)H^2a^2(\tau')\frac{\Delta x_\rho\Delta x_\sigma}{y^{3-\varepsilon}}\notag\\
&\hspace{1em}+H^2a^2(\tau')\frac{\Delta x_\rho\Delta x_\sigma}{y^2}
+2Ha(\tau')\frac{\Delta x_\rho\delta_\sigma^{\ 0}+\Delta x_\sigma\delta_\rho^{\ 0}}{y^{3-\varepsilon}}\notag\\
&\hspace{1em}+Ha(\tau')\frac{\Delta x_\rho\delta_\sigma^{\ 0}+\Delta x_\sigma\delta_\rho^{\ 0}}{y^2}
+\frac{\delta_\rho^{\ 0}\delta_\sigma^{\ 0}}{y^{2-\varepsilon}}\big]. \notag
\end{align}
Note that we have only to evaluate (\ref{int1}) up to $\mathcal{O}(\varepsilon^0)$. 
By substituting (\ref{G1}) and (\ref{G2}) to (\ref{int1}), 
\begin{align}
\parbox{\cloverdl}{\usebox{\cloverd}}
\simeq i\frac{2^{2D-3}H^{4D-4}}{(4\pi)^{2D}}(D-1)\Gamma^2(D-1)a^2(\tau)\int d^{4-\varepsilon}x'\ a^{4-\varepsilon}(\tau')\log a(\tau')\sum_{m=1}^{6}F^m_{\rho\sigma}, 
\label{int2}\end{align}
where the integrands are as follows
\begin{align}
F^1_{\rho\sigma}\equiv 4H^2a^2(\tau')[\frac{\Delta x_\rho\Delta x_\sigma}{y^{4-\varepsilon}_{++}}-\frac{\Delta x_\rho\Delta x_\sigma}{y^{4-\varepsilon}_{++}}], 
\label{F1}\end{align}
\begin{align}
F^2_{\rho\sigma}\equiv (4-\varepsilon)H^2a^2(\tau')[\frac{\Delta x_\rho\Delta x_\sigma}{y^{3-\varepsilon}_{++}}-\frac{\Delta x_\rho\Delta x_\sigma}{y^{3-\varepsilon}_{+-}}], 
\label{F2}\end{align}
\begin{align}
F^3_{\rho\sigma}\equiv H^2a^2(\tau')[\frac{\Delta x_\rho\Delta x_\sigma}{y^2_{++}}-\frac{\Delta x_\rho\Delta x_\sigma}{y^2_{+-}}], 
\label{F3}\end{align}
\begin{align}
F^4_{\rho\sigma}\equiv 2Ha(\tau')[\frac{\Delta x_\rho\delta_\sigma^{\ 0}+\Delta x_\sigma\delta_\rho^{\ 0}}{y^{3-\varepsilon}_{++}}
-\frac{\Delta x_\rho\delta_\sigma^{\ 0}+\Delta x_\sigma\delta_\rho^{\ 0}}{y^{3-\varepsilon}_{+-}}], 
\label{F4}\end{align}
\begin{align}
F^5_{\rho\sigma}\equiv Ha(\tau')[\frac{\Delta x_\rho\delta_\sigma^{\ 0}+\Delta x_\sigma\delta_\rho^{\ 0}}{y^2_{++}}
-\frac{\Delta x_\rho\delta_\sigma^{\ 0}+\Delta x_\sigma\delta_\rho^{\ 0}}{y^2_{+-}}], 
\label{F5}\end{align}
\begin{align}
F^6_{\rho\sigma}\equiv [\frac{\delta_\rho^{\ 0}\delta_\sigma^{\ 0}}{y^{2-\varepsilon}_{++}}-\frac{\delta_\rho^{\ 0}\delta_\sigma^{\ 0}}{y^{2-\varepsilon}_{+-}}]. 
\label{F6}\end{align}

First, we calculate the integral of $F_{\rho\sigma}^1$
\begin{align}
&\int d^{4-\varepsilon}x'\ a^{4-\varepsilon}(\tau')\log a(\tau')F^1_{\rho\sigma}\label{F1_1}\\
=&\ 4H^{-6+2\varepsilon}a^{-4+\varepsilon}(\tau)\int d^{4-\varepsilon}x'\ a^2(\tau')\log a(\tau')
[\frac{\Delta x_\rho\Delta x_\sigma}{\Delta x^{8-2\varepsilon}_{++}}
-\frac{\Delta x_\rho\Delta x_\sigma}{\Delta x^{8-2\varepsilon}_{+-}}]. \notag
\end{align}
The integrand is written as follows
\begin{align}
\frac{\Delta x_\rho\Delta x_\sigma}{\Delta x^{8-2\varepsilon}}
=\frac{-1}{(6-2\varepsilon)(4-2\varepsilon)(2-2\varepsilon)}(\partial_\rho\partial_\sigma+\frac{\eta_{\rho\sigma}\partial^2}{2-\varepsilon})
\frac{\partial^2}{\varepsilon}\frac{1}{\Delta x^{2-2\varepsilon}}, 
\end{align}
where we abbreviate the indexes $++$, $+-$ because the above identities work out in the both case. 
By using this identity, (\ref{F1_1}) is
\begin{align}
&\int d^{4-\varepsilon}x'\ a^{4-\varepsilon}(\tau')\log a(\tau')F^1_{\rho\sigma}\\
=&\ \frac{-4H^{-6}}{(6-2\varepsilon)(4-2\varepsilon)(2-2\varepsilon)}
(-\delta_\rho^{\ 0}\delta_\sigma^{\ 0}+\frac{\eta_{\rho\sigma}}{2-\varepsilon})\notag\\
&\times H^{2\varepsilon}a^{4-\varepsilon}(\tau)\partial^2\int d^{4-\varepsilon}x'\ a^2(\tau')\log a(\tau')
\frac{\partial^2}{\varepsilon}[\frac{1}{\Delta x^{2-2\varepsilon}_{++}}-\frac{1}{\Delta x^{2-2\varepsilon}_{+-}}]. \notag
\end{align}
Note that the differential operator outside the integral is equal to the time derivative 
$\partial_\rho\to\delta_\rho^{\ 0}\partial_0$, $\partial^2\to-\partial^2_0$.   
The second order differentials of $1/\Delta x^{2-\varepsilon}$ are
\begin{align}
\partial^2\frac{1}{\Delta x^{2-\varepsilon}_{++}}=\frac{4i\pi^{2-\frac{\varepsilon}{2}}}{\Gamma(1-\frac{\varepsilon}{2})}\delta^{(D)}(x-x'),\hspace{1em}
\partial^2\frac{1}{\Delta x^{2-\varepsilon}_{+-}}=0. 
\label{2d}\end{align}
By using (\ref{2d}), we extract the UV divergent part
\begin{align}
\frac{\partial^2}{\varepsilon}\frac{1}{\Delta x^{2-2\varepsilon}_{++}}
&=\frac{\partial^2}{\varepsilon}(\frac{1}{\Delta x^{2-2\varepsilon}_{++}}-\frac{\mu^{-\varepsilon}}{\Delta x^{2-\varepsilon}_{++}})
+\frac{4i\pi^{2-\frac{\varepsilon}{2}}\mu^{-\varepsilon}}{\Gamma(1-\frac{\varepsilon}{2})}\delta^{(D)}(x-x')\label{UVextract}\\
&=\frac{\partial^2}{2}\frac{\log (\mu^2\Delta x^2)}{\Delta x^2_{++}}
+\frac{4i\pi^{2-\frac{\varepsilon}{2}}\mu^{-\varepsilon}}{\Gamma(1-\frac{\varepsilon}{2})}\delta^{(D)}(x-x'), \notag\\
\frac{\partial^2}{\varepsilon}\frac{1}{\Delta x^{2-2\varepsilon}_{+-}}
&=\frac{\partial^2}{\varepsilon}(\frac{1}{\Delta x^{2-2\varepsilon}_{+-}}-\frac{\mu^{-\varepsilon}}{\Delta x^{2-\varepsilon}_{+-}})
=\frac{\partial^2}{2}\frac{\log (\mu^2\Delta x^2)}{\Delta x^2_{+-}}, \notag
\end{align}
where we introduce the mass parameter $\mu$ to correct the dimension. 
From (\ref{log1}), (\ref{log2}) and (\ref{UVextract}), 
\begin{align}
&\ H^{2\varepsilon}a^{4-\varepsilon}(\tau)\partial^2\int d^{4-\varepsilon}x'\ a^2(\tau')\log a(\tau')
\frac{\partial^2}{\varepsilon}[\frac{1}{\Delta x^{2-2\varepsilon}_{++}}-\frac{1}{\Delta x^{2-2\varepsilon}_{+-}}]\\
=&\ H^{2\varepsilon}a^{4-\varepsilon}(\tau)\partial^2
\Big\{\frac{4i\pi^{2-\frac{\varepsilon}{2}}\mu^{-\varepsilon}}{\Gamma(1-\frac{\varepsilon}{2})}a^2(\tau)\log a(\tau)\notag\\
&\hspace{6em}+i\pi^2\partial_0^4\int^\tau_{-\frac{1}{H}}d\tau'\ a^2(\tau')\log a(\tau')\int^{\Delta \tau}_0r^2dr\ 
\big(\log \mu^2(\Delta \tau^2-r^2)-1\big)\Big\}\notag\\
=&\ H^{2\varepsilon}a^{4-\varepsilon}(\tau)\partial^2
\Big\{\frac{4i\pi^{2-\frac{\varepsilon}{2}}\mu^{-\varepsilon}}{\Gamma(1-\frac{\varepsilon}{2})\varepsilon}a^2(\tau)\log a(\tau)
+4i\pi^2\partial_0\int^\tau_{-\frac{1}{H}}d\tau'\ a^2(\tau')\log a(\tau')\log 2\mu\Delta\tau \Big\}\notag\\
=&\ 4i\pi^2a^{4-\varepsilon}(\tau)\partial^2
\Big\{\frac{\pi^{-\frac{\varepsilon}{2}}\mu^{-\varepsilon}H^{2\varepsilon}}{\Gamma(1-\frac{\varepsilon}{2})\varepsilon}a^2(\tau)\log a(\tau)
+a^2(\tau)\log a(\tau)\log \frac{2\mu}{H}-a^2(\tau)\log^2 a(\tau)\notag\\
&\hspace{6em}-a^2(\tau)\frac{\partial}{\partial a(\tau)}\int^{a(\tau)}_1da(\tau')\ \log a(\tau')\sum_{n=1}^\infty \frac{a^n(\tau')}{na^n(\tau)}\Big\}\notag\\
\simeq&-4i\pi^2H^2
\Big\{6(\frac{\pi^{-\frac{\varepsilon}{2}}\mu^{-\varepsilon}H^{2\varepsilon}}{\Gamma(1-\frac{\varepsilon}{2})\varepsilon}+\log\frac{2\mu}{H})\log a(\tau)
-11\log a(\tau)\Big\}. 
\end{align}
So the integral of $F^1_{\rho\sigma}$ is
\begin{align}
&\int d^{4-\varepsilon}x'\ a^{4-\varepsilon}(\tau')\log a(\tau')F^1_{\rho\sigma}\label{F1_3}\\
\simeq&\ \eta_{\rho\sigma}\times 4i\pi^2H^{-4}
\Big\{\frac{1}{4}(\frac{\pi^{-\frac{\varepsilon}{2}}\mu^{-\varepsilon}H^{2\varepsilon}}{\Gamma(1-\frac{\varepsilon}{2})\varepsilon}+\log\frac{2\mu}{H})\log a(\tau)
+\frac{1}{8}\log a(\tau)\Big\}\notag\\
&+\delta_\rho^{\ 0}\delta_\sigma^{\ 0}\times 4i\pi^2H^{-4}
\Big\{-\frac{1}{2}\big(\frac{\pi^{-\frac{\varepsilon}{2}}\mu^{-\varepsilon}H^{2\varepsilon}}{\Gamma(1-\frac{\varepsilon}{2})\varepsilon}+\log\frac{2\mu}{H}\big)
\log a(\tau)\Big\}. \notag
\end{align}
In a similar way, the other integrals are
\begin{align}
&\int d^{4-\varepsilon}x'\ a^{4-\varepsilon}(\tau')\log a(\tau')F^2_{\rho\sigma}\label{F2_1}\\
\simeq&\ \eta_{\rho\sigma}\times 4i\pi^2H^{-4}
\Big\{-\frac{1}{2}\big(\frac{\pi^{-\frac{\varepsilon}{2}}\mu^{-\varepsilon}H^{2\varepsilon}}{\Gamma(1-\frac{\varepsilon}{2})\varepsilon}+\log\frac{2\mu}{H}\big)\log a(\tau)
+\frac{1}{8}\log a(\tau)\Big\}\notag\\
&+\delta_\rho^{\ 0}\delta_\sigma^{\ 0}\times 4i\pi^2H^{-4}
\Big\{-\frac{1}{2}\log a(\tau)\Big\}, \notag
\end{align}
\begin{align}
&\int d^{4}x'\ a^{4}(\tau')\log a(\tau')F^3_{\rho\sigma}\label{F3_1}\\
\simeq&\ \eta_{\rho\sigma}\times 4i\pi^2H^{-4}
\Big\{-\frac{1}{12}\log a(\tau)\Big\}
+\delta_\rho^{\ 0}\delta_\sigma^{\ 0}\times 4i\pi^2H^{-4}
\Big\{-\frac{1}{6}\log a(\tau)\Big\}, \notag
\end{align}
\begin{align}
&\int d^{4-\varepsilon}x'\ a^{4-\varepsilon}(\tau')\log a(\tau')F^4_{\rho\sigma}\label{F4_1}\\
\simeq&\ \delta_\rho^{\ 0}\delta_\sigma^{\ 0}\times 4i\pi^2H^{-4}
\Big\{\big(\frac{\pi^{-\frac{\varepsilon}{2}}\mu^{-\varepsilon}H^{2\varepsilon}}{\Gamma(1-\frac{\varepsilon}{2})\varepsilon}+\log\frac{2\mu}{H}\big)\log a(\tau)
\Big\}, \notag
\end{align}
\begin{align}
&\int d^{4}x'\ a^{4}(\tau')\log a(\tau')F^5_{\rho\sigma}\label{F5_1}\\
\simeq&\ \delta_\rho^{\ 0}\delta_\sigma^{\ 0}\times 4i\pi^2H^{-4}
\Big\{\frac{1}{2}\log a(\tau)
\Big\}, \notag
\end{align}
\begin{align}
&\int d^{4-\varepsilon}x'\ a^{4-\varepsilon}(\tau')\log a(\tau')F^6_{\rho\sigma}\label{F6_1}\\
\simeq&\ \delta_\rho^{\ 0}\delta_\sigma^{\ 0}\times 4i\pi^2H^{-4}
\Big\{-\frac{1}{2}\big(\frac{\pi^{-\frac{\varepsilon}{2}}\mu^{-\varepsilon}H^{2\varepsilon}}{\Gamma(1-\frac{\varepsilon}{2})\varepsilon}
+\log\frac{2\mu}{H}\big)\log a(\tau)\Big\}, \notag
\end{align}
From (\ref{int2}) and (\ref{F1_3})-(\ref{F6_1}), 
\begin{align}
\parbox{\cloverdl}{\usebox{\cloverd}}
\simeq&\ g_{\rho\sigma}\frac{2^{2D-3}\pi^2H^{4D-8}}{(4\pi)^{2D}}(D-1)\Gamma^2(D-1) \label{int3}\\
&\times\Big\{\big(\frac{\pi^{-\frac{\varepsilon}{2}}\mu^{-\varepsilon}H^{2\varepsilon}}{\Gamma(1-\frac{\varepsilon}{2})\varepsilon}
+\log\frac{2\mu}{H}\big)\log a(\tau) -\frac{2}{3}\log a(\tau)\Big\} 
+a^2(\tau)\delta_\rho^{\ 0}\delta_\sigma^{\ 0}\frac{H^8}{2^8\pi^6}\log a(\tau). \notag
\end{align}
From (\ref{int0}) and (\ref{int3}), 
\begin{align}
\parbox{\cloverdl}{\usebox{\cloverd}}-\parbox{\cloverbl}{\usebox{\cloverb}}
\simeq&\ g_{\rho\sigma}\frac{2^{2D-3}\pi^2H^{4D-8}}{(4\pi)^{2D}}(D-1)\Gamma^2(D-1) \\
&\times\Big\{\big(\frac{\pi^{-\frac{\varepsilon}{2}}\mu^{-\varepsilon}H^{2\varepsilon}}{\Gamma(1-\frac{\varepsilon}{2})\varepsilon}
+\log\frac{2\mu}{H}\big)\log a(\tau) -\frac{2}{3}\log a(\tau)\Big\}. \notag
\end{align}
As a result, the contribution from the two diagrams is
\begin{align}
\langle T_{\mu\nu}\rangle|_{g^4}\simeq&\ (\delta_\mu^{\ \rho}\delta_\nu^{\ \sigma}-\frac{1}{2}g_{\mu\nu}g^{\rho\sigma})
\times-\frac{g^4}{2}D^2R\Big[\parbox{\cloverdl}{\usebox{\cloverd}}-\parbox{\cloverbl}{\usebox{\cloverb}}\Big]\\
\simeq&\ g_{\mu\nu}g^4D^2R\frac{2^{2D-4}\pi^2H^{4D-8}}{(4\pi)^{2D}}(D-1)\Gamma^2(D-1) \notag\\
&\times\Big\{\big(\frac{\pi^{-\frac{\varepsilon}{2}}\mu^{-\varepsilon}H^{2\varepsilon}}{\Gamma(1-\frac{\varepsilon}{2})\varepsilon}
+\log\frac{2\mu}{H}\big)\log a(\tau) -\frac{7}{6}\log a(\tau)\Big\} \notag\\
=&\ g_{\mu\nu}g^4D^2R\frac{(D-1)(D-2)}{2}\frac{H^{3D-4}}{(4\pi)^\frac{3D}{2}}\frac{\Gamma^2(D-1)}{\Gamma(\frac{D}{2})}
\Big\{\frac{1}{\varepsilon}\log a(\tau)-\frac{7}{6}\log a(\tau)\Big\}. \notag
\end{align}


\end{document}